\pdfoutput=1

\documentclass[prd,aps,twocolumn,preprintnumbers,amssymb,nobibnotes,nofootinbib,longbibliography,superscriptaddress,10pt]{revtex4-1}

\usepackage[left=1in, right=1in, top=1in, bottom=1in]{geometry}   

\usepackage[utf8]{inputenc}
\usepackage[english]{babel}  
\usepackage{hyperref}
\hypersetup{
    colorlinks=true,
    allcolors={blue!60!black}
}

\usepackage{bbold}
\usepackage{graphicx,epsf}
\usepackage{amsmath,amsfonts,amssymb,amsbsy,mathrsfs}

\usepackage{tikz}
\usetikzlibrary{decorations.pathmorphing}
\usetikzlibrary{decorations.markings}
\usetikzlibrary{positioning, shapes, arrows}
\usetikzlibrary{calc}
\usepackage[compat=1.0.0]{tikz-feynman}

\tikzset{
quark/.style={postaction={decorate}, decoration={markings}},
scalar/.style={dashed,postaction={decorate}, decoration={markings,mark=at position .5 with {\arrow[#1]{latex}}}},
gluon/.style={decorate,decoration={coil,amplitude=3pt, segment length=4.7pt, pre length=.01cm, post length=.01cm}},
gluont/.style={decorate,decoration={coil,amplitude=3pt, segment length=3.50pt, pre length=.01cm, post length=.01cm}},
}

\usepackage{caption}
\usepackage{subcaption}
\usepackage{physics}

\usepackage{color,xcolor}

\usepackage{array,multirow}
\usepackage{stackrel}

\usepackage{ifdraft}
\usepackage[obeyFinal]{easy-todo}

\usepackage{placeins}
\usepackage{slashed,shuffle} 
\usepackage{adjustbox}

\usepackage{comment}
\usepackage[normalem]{ulem}

\usepackage{slashed}
\usepackage{changepage}

\usepackage{cleveref}
\usepackage{pifont}

\DeclareUnicodeCharacter{27E8}{{\langle}}
\DeclareUnicodeCharacter{27E9}{{\rangle}}

\newcolumntype{C}[1]{>{\centering\arraybackslash}m{#1}}
\newcolumntype{M}{>{\centering\arraybackslash$}l<{$}}

\newcommand{\eps}{\epsilon}
\def\trFive{{\rm tr}_5}

\newcommand{\MSbar}{\overline{\text{MS}}}

\newcommand{\ii}{\mathrm{i}}

\newcommand{\red}[1]{\textcolor{red!50!black}{#1}}

\begin{document}

\title{Double-Virtual NNLO QCD Corrections for Five-Parton Scattering:\protect\\ The Quark Channels}
\preprint{PSI-PR-24-04, ZU-TH 78/23}


%
\author{Giuseppe~De~Laurentis}
\affiliation{Higgs Centre for Theoretical Physics, University of Edinburgh, Edinburgh, EH9 3FD, United Kingdom}
\author{Harald~Ita}
\affiliation{Paul Scherrer Institut, Forschungsstrasse 111, 5232 Villigen, Switzerland}
\affiliation{ICS, University of Zurich, Winterthurerstrasse 190, Zurich, Switzerland}
\author{Vasily~Sotnikov}
\affiliation{Physik-Institut, University of Zurich, Winterthurerstrasse 190, 8057 Zurich, Switzerland}
%

\begin{abstract}
\begin{adjustwidth}{-10mm}{10mm} 
\begin{quotation}
We complete the computation of two-loop helicity amplitudes required to
obtain next-to-next-to-leading order QCD corrections 
for three-jet production at hadron colliders, including all
contributions beyond the leading-color approximation.
The analytic expressions are reconstructed from finite-field samples obtained
with the numerical unitarity method.
We find that the reconstruction is significantly facilitated by exploiting the
overlaps between rational coefficient functions of quark and gluon processes,
and we display their compact generating sets in the appendix of the paper.
We implement our results in a public code, and demonstrate its suitability
for phenomenological applications.
\end{quotation}
\end{adjustwidth}
\end{abstract}

\maketitle

\section{Introduction}\label{sec:introduction}

Scattering processes that produce multiple jets in the final states are 
abundant at the Large Hadron Collider. In the next decades, precise theory
predictions suitable for analyzing the expected large event samples will be
crucial for advancing our understanding of high-energy interactions.
Historically, pure-QCD predictions aimed at multi-jet production have also
played an important role in the discovery of new structures and methods in
field-theory, which inspire neighboring fields. 

The goal of this work are predictions for the five-parton processes in QCD,
which have been an active field of research for many years.  
Tree-level five-point amplitudes in QCD were
derived a long time ago in 1985 \cite{Parke:1985pn}.  About a decade later the
one-loop amplitudes were obtained \cite{Bern:1993mq,Kunszt:1994nq,Bern:1994fz}.  
The increase in analytic and algebraic complexity of two-loop computations
required significant theoretical and technical developments over the following
20 years. This led to the five-point two-loop amplitudes for the all-plus helicity configuration
to be first computed numerically
\cite{Badger:2013gxa} and later in analytic form
\cite{Gehrmann:2015bfy,Dunbar:2016aux}. By now, the five-parton 
two-loop amplitudes with all helicity configurations are known in the leading-color approximation  \cite{Badger:2017jhb,Abreu:2017hqn,Badger:2018gip,Abreu:2018jgq,Badger:2018enw,Abreu:2018zmy,Abreu:2019odu,Abreu:2021oya}.

These results opened the door for the first computation of next-to-next-to-leading order (NNLO) QCD
predictions for three-jet production \cite{Czakon:2021mjy} (see also
\cite{Chen:2022ktf}), and the follow-up measurements of the strong coupling
constant at high scales \cite{Czakon:2021mjy,ATLAS:2023tgo,Alvarez:2023fhi}.
The double-virtual corrections \cite{Abreu:2021oya}, contributing on average about $10\%$ \cite{Czakon:2021mjy},
have been included in the leading color approximation in these works.
This highlights the potential importance of including
subleading-color effects.

In the preceding publication \cite{DeLaurentis:2023nss} we have obtained the analytic expressions for the five-point gluon
amplitudes including all color
contributions.  Computing the mixed quark and gluon channels is the central
goal of this work, which completes the two-loop five-parton scattering amplitudes.

The computation of the non-planar five-point amplitudes is a multi-layered
challenge, due to its analytic and combinatorial complexity.  We rely on a
number of recent results and methods. We use the massless five-point Feynman
integrals
\cite{Papadopoulos:2015jft,Gehrmann:2018yef,Abreu:2018aqd,Chicherin:2018old,Chicherin:2020oor},
rely on the geometric methods for obtaining integration-by-parts identities
\cite{Gluza:2010ws,vonManteuffel:2014ixa,Ita:2015tya,Larsen:2015ped,Bern:2017gdk}, and we
employ analytic reconstruction methods
\cite{Peraro:2016wsq,Klappert:2020nbg,Magerya:2022hvj,Belitsky:2023qho,DeLaurentis:2022otd,Badger:2021imn,Abreu:2021asb,Liu:2023cgs,DeLaurentis:2019bjh,DeLaurentis:2020qle,Campbell:2022qpq}. 
In our computation we closely follow our recent work \cite{DeLaurentis:2023nss}.
We apply the numerical unitarity
method~\cite{Ita:2015tya,Abreu:2017xsl,Abreu:2017hqn,Abreu:2020xvt,Abreu:2023bdp}
to compute numerical values for the scattering amplitudes, which we use to
reconstruct analytic results.
The computation relies on the spinor-helicity formalism, which in turn yields 
very compact expressions as displayed in the appendices.
In order to reduce the computational load,
we develop an efficient way to obtain a large portion of quark amplitudes from gluon
amplitudes. In fact, we rescale gluon amplitudes in a way 
reminiscent of supersymmetry Ward 
identities \cite{Grisaru:1977px,Grisaru:1976vm,Parke:1985pn}. The remaining parts of the amplitude are
obtained applying a variant of a recent analytic reconstruction technique \cite{Abreu:2023bdp}.

The analytic results for amplitudes are provided in
ancillary files, suitable for future theoretical and phenomenological studies. 
In particular, we provide a \texttt{C++} library for fast and stable numerical evaluation of our analytic results.   
Together with the gluon channel, presented in the first part of this work \cite{DeLaurentis:2023nss},
these results will provide crucial input for NNLO predictions for
three-jet and for N\textsuperscript{3}LO di-jet production in hadron collisions. 

\vspace{1em} \textbf{Note added:} while this work was in preparation, partially
overlapping results were reported \cite{Agarwal:2023suw}.
We thank the authors of ref.~\cite{Agarwal:2023suw} for correspondence on
numerical comparison between our results.

\section{Notation and Conventions}\label{sec:notation}

\subsection{Helicity amplitudes}
\label{sec:helAmpl}

We consider the channels of the five-parton scattering in QCD that involve external quarks. We
associate indices 1 and 2 to the initial state partons and 3, 4 and 5 to final
states. There are three channels with a pair of quark and anti-quark,
\begin{align} \label{eq:2q3g}
  \bar{u}_{-p_1}^{-h_1}+u_{-p_2}^{-h_2}\to g_{p_3}^{h_3}&+g_{p_4}^{h_4}+g_{p_5}^{h_5}\,, \\
  \bar{u}_{-p_1}^{-h_1}+g_{-p_2}^{-h_2}\to \bar{u}_{p_3}^{h_3}&+g_{p_4}^{h_4}+g_{p_5}^{h_5}\,, \\ 
  g_{-p_1}^{-h_1}+g_{-p_2}^{-h_2}\to u_{p_3}^{h_3}&+{\bar u}_{p_4}^{h_4}+g_{p_5}^{h_5}\,.
\end{align}
Without loss of generality we denote the quarks with $u$ for up-type
quark. There are four quark channels with distinct quark flavors,
chosen to be up-type and down-type quarks,
\begin{align} \label{eq:4q1g}
  \bar{u}_{-p_1}^{-h_1}+u_{-p_2}^{-h_2}\to d_{p_3}^{h_3}&+\bar{d}_{p_4}^{h_4}+g_{p_5}^{h_5}\,, \\
  \bar{u}_{-p_1}^{-h_1}+d_{-p_2}^{-h_2}\to d_{p_3}^{h_3}&+\bar{u}_{p_4}^{h_4}+g_{p_5}^{h_5}\,, \\
  \bar{u}_{-p_1}^{-h_1}+\bar{d}_{-p_2}^{-h_2}\to \bar{d}_{p_3}^{h_3}&+\bar{u}_{p_4}^{h_4}+g_{p_5}^{h_5}\,, \\
  \bar{u}_{-p_1}^{-h_1}+g_{-p_2}^{-h_2}\to \bar{u}_{p_3}^{h_3}&+d_{p_4}^{h_4}+\bar{d}_{p_5}^{h_5}\,.
\end{align}
All other channels involving distinct quark flavors are related to the above by charge
conjugation and permutation of labels. The three channels with four identical quark flavors 
are obtained as linear combination of the distinct quark ones,
\begin{align}\label{eq:4q1g_identical}
  \bar{u}_{-p_1}^{-h_1}+&u_{-p_2}^{-h_2}\to u_{p_3}^{h_3}+\bar{u}_{p_4}^{h_4}+g_{p_5}^{h_5}
= \\
  &\big(\bar{u}_{-p_1}^{-h_1}+u_{-p_2}^{-h_2}\to d_{p_3}^{h_3}+\bar{d}_{p_4}^{h_4}+g_{p_5}^{h_5}\big) - \nonumber\\
  &\big(\bar{u}_{-p_1}^{-h_1}+d_{-p_2}^{-h_2}\to d_{p_3}^{h_3}+\bar{u}_{p_4}^{h_4}+g_{p_5}^{h_5}\big)\,,\nonumber\\
  \bar{u}_{-p_1}^{-h_1}+&\bar{u}_{-p_2}^{-h_2}\to \bar{u}_{p_3}^{h_3}+\bar{u}_{p_4}^{h_4}+g_{p_5}^{h_5}
= \\
  &\big(\bar{u}_{-p_1}^{-h_1}+\bar{u}_{-p_2}^{-h_2}\to \bar{d}_{p_3}^{h_3}+\bar{d}_{p_4}^{h_4}+g_{p_5}^{h_5}\big) - \nonumber\\
  &\big(\bar{u}_{-p_1}^{-h_1}+\bar{d}_{-p_2}^{-h_2}\to \bar{d}_{p_3}^{h_3}+\bar{u}_{p_4}^{h_4}+g_{p_5}^{h_5}\big)\,,\nonumber\\
  \bar{u}_{-p_1}^{-h_1}+&g_{-p_2}^{-h_2}\to \bar{u}_{p_3}^{h_3}+u_{p_4}^{h_4}+\bar{u}_{p_5}^{h_5}
= \\
  &\big(\bar{u}_{-p_1}^{-h_1}+g_{-p_2}^{-h_2}\to \bar{u}_{p_3}^{h_3}+d_{p_4}^{h_4}+\bar{d}_{p_5}^{h_5}\big) - \nonumber\\
  &\big(\bar{u}_{-p_1}^{-h_1}+g_{-p_2}^{-h_2}\to \bar{d}_{p_3}^{h_3}+d_{p_4}^{h_4}+\bar{u}_{p_5}^{h_5}\big)\,.\nonumber
\end{align}

Representative diagrams for the
two-loop contributions are collected in table
\ref{tab:diagram_2q3g} and 
\ref{tab:diagram_4q1g}.

Throughout this article we use two interchangeable notations to specify the particles'
helicities. We either give signs $\pm$ to label positive/negative helicities or,
equivalently, the (half-)integers $h=\pm 1/2$ and $h=\pm 1$ to provide a better distinction between the quark and gluon
helicities, respectively.

Whenever we use the arrow notation ($\to$), the first two particles are to be
understood as crossed to be incoming, such that their quantum numbers and
momenta are given in incoming convention, while the final states are
understood in out-going convention. If no arrow ($\to$) is used we consider the
all-outgoing convention.

\begin{table}
\begin{subfigure}{0.48\linewidth}
\begin{tikzpicture}[scale=0.6]
\node (g1) at (-1,1){};
\node (g2) at (-1,-3){};
\node (g3) at (5,1){};
\node (g4) at (5,-3){};
\node (g5) at (3.5,-1){};
\coordinate (a1) at (0,0);
\coordinate (a2) at (0,-2);
\coordinate (b1) at (2,0);
\coordinate (b3) at (2,-1);
\coordinate (b2) at (2,-2);
\coordinate (c1) at (4,0);
\coordinate (c2) at (4,-2);
\draw [quark] (g1) -- (a1) -- (a2) -- (g2);
\draw [gluon] (c2) -- (g4);
\draw [gluon] (c1) -- (g3);
\draw [gluon] (a1) -- (b1) -- (c1);
\draw [gluon] (a2) -- (b2) -- (c2);
\draw [gluon] (b1) -- (b2);
\draw [gluon] (b3) -- (g5);
\draw [gluon] (c1) -- (c2);
\end{tikzpicture}
\end{subfigure}
~
\begin{subfigure}{0.48\linewidth}
\begin{tikzpicture}[scale=0.6]
\node (g1) at (-1,1){};
\node (g2) at (-1,-3){};
\node (g3) at (5,1){};
\node (g4) at (5,-3){};
\node (g5) at (3.5,-1){};
\coordinate (a1) at (0,0);
\coordinate (a2) at (0,-2);
\coordinate (b1) at (2,0);
\coordinate (b3) at (2,-1);
\coordinate (b2) at (2,-2);
\coordinate (c1) at (4,0);
\coordinate (c2) at (4,-2);
\draw [quark] (g1) -- (a1) -- (a2) -- (g2);
\draw [gluon] (c1) -- (g3);
\draw [gluon] (a1) -- (b1);
\draw [gluon] (c2) -- (g4);
\draw [gluon] (a2) -- (b2);
\draw [quark] (b1) -- (b2) -- (c2) -- (c1) -- (b1);
\draw [gluon] (b3) -- (g5);
\end{tikzpicture}
\end{subfigure}
\begin{subfigure}{0.48\linewidth}
\begin{tikzpicture}[scale=0.6]
\node (g1) at (-1.4,0){};
\node (g2) at (-1.4,-2){};
\node (g3) at (3.5,1.3){};
\node (g4) at (3.5,-3.3){};
\node (g5) at (5.9,-1){};
\coordinate (a1) at (-0.4,-1);
\coordinate (a2) at (0.4,-1);
\coordinate (b1) at (1.6,-1);
\coordinate (b2) at (2.4,-1);
\coordinate (c1) at (3.5,0);
\coordinate (c2) at (3.5,-2);
\coordinate (d1) at (4.6,-1);
\draw[quark] (b1) arc (0:360:0.6);
\draw [gluont] (a1) -- (a2);
\draw [gluont] (b1) -- (b2);
\draw [quark] (g1) -- (a1);
\draw [quark] (g2) -- (a1);
\draw [gluon] (g3) -- (c1);
\draw [gluon] (g4) -- (c2);
\draw [gluont] (g5) -- (d1);
\draw [quark] (b2) -- (c1) -- (d1) -- (c2) -- (b2); 
\end{tikzpicture}
\end{subfigure}
\caption{ Representative Feynman diagrams for two-quark three-gluon amplitudes, contributing at order $N^0_f$, $N^1_f$ and $N^2_f$.}
\label{tab:diagram_2q3g}
\end{table}
\begin{table}
\begin{subfigure}{0.48\linewidth}
\begin{tikzpicture}[scale=0.6]
\node (g1) at (-1,1){};
\node (g2) at (-1,-3){};
\node (g3) at (5,1){};
\node (g4) at (5,-3){};
\node (g5) at (3.5,-1){};
\coordinate (a1) at (0,0);
\coordinate (a2) at (0,-2);
\coordinate (b1) at (2,0);
\coordinate (b3) at (2,-1);
\coordinate (b2) at (2,-2);
\coordinate (c1) at (4,0);
\coordinate (c2) at (4,-2);
\draw [quark] (g1) -- (a1) -- (a2) -- (g2);
\draw [quark] (g3) -- (c1) -- (c2) -- (g4);
\draw [gluon] (a1) -- (b1) -- (c1);
\draw [gluon] (a2) -- (b2) -- (c2);
\draw [gluon] (b1) -- (b2);
\draw [gluon] (b3) -- (g5);
\end{tikzpicture}
\end{subfigure}
~
\begin{subfigure}{0.48\linewidth}
\begin{tikzpicture}[scale=0.6]
\node (g1) at (-1.2,0){};
\node (g2) at (-1.2,-2){};
\node (g3) at (5,1){};
\node (g4) at (5,-3){};
\node (g5) at (3.5,-1){};
\coordinate (a1) at (-0.4,-1);
\coordinate (a2) at (0.4,-1);
\coordinate (b1) at (2,0);
\coordinate (b3) at (2,-1);
\coordinate (b2) at (2,-2);
\coordinate (c1) at (4,0);
\coordinate (c2) at (4,-2);
\draw [quark] (g1) -- (a1) -- (g2);
\draw [quark] (g3) -- (c1) -- (c2) -- (g4);
\draw [gluont] (a1) -- (a2);
\draw [quark] (a2) -- (b1) -- (b2) -- (a2);
\draw [gluon] (b2) -- (c2);
\draw [gluon] (b1) -- (c1);
\draw [gluon] (b3) -- (g5);
\end{tikzpicture}
\end{subfigure}

\begin{subfigure}{0.48\linewidth}
\begin{tikzpicture}[scale=0.6]
\node (g1) at (-1.4,0){};
\node (g2) at (-1.4,-2){};
\node (g3) at (5.2,1.3){};
\node (g4) at (5.8,-1.6){};
\node (g5) at (5.8,-3.2){};
\coordinate (a1) at (-0.4,-1);
\coordinate (a2) at (0.4,-1);
\coordinate (b1) at (1.6,-1);
\coordinate (b2) at (2.4,-1);
\coordinate (c1) at (4,0);
\coordinate (c2) at (4,-2);
\coordinate (d1) at (4.8,-2.4);
\draw[quark] (b1) arc (0:360:0.6);
\draw [gluont] (a1) -- (a2);
\draw [gluont] (b1) -- (b2);
\draw [quark] (g1) -- (a1);
\draw [quark] (g2) -- (a1);
\draw [gluon] (g3) -- (c1);
\draw [quark] (g4) -- (d1) -- (g5);
\draw [gluont] (c2) -- (d1);
\draw [quark] (b2) -- (c1) -- (c2) -- (b2); 
\end{tikzpicture}
\end{subfigure}
\caption{ Representative Feynman diagrams for four-quark one-gluon amplitudes, contributing at order $N^0_f$, $N^1_f$ and $N^2_f$.}
\label{tab:diagram_4q1g}
\end{table}

\subsection{Kinematics and permutation groups}
\label{sec:kinematics}

We consider the scattering of five massless particles using the same conventions as in ref.~\cite{DeLaurentis:2023nss}. 
The kinematic is defined with five Mandelstam invariants $\{s_{12},s_{23},s_{34},s_{45},s_{15}\}$, together with 
the parity-odd contraction $\trFive = \trace(\gamma^5\slashed{p}_1\slashed{p}_2\slashed{p}_3\slashed{p}_4)$.

The particles' helicity states are specified using
two-component spinors, $\lambda_{i}^{\alpha}$ and $\tilde\lambda_{i}^{\dot\alpha}$, with $i \in \{1, \dots, 5\}$. 
We define spinor brackets as the contractions,
\begin{equation}
  \langle i j \rangle = \lambda^\alpha_i \lambda_{j, \alpha} \quad \text{and} \quad [ij] = \tilde\lambda_{i,\dot\alpha}\tilde\lambda_j^{\dot\alpha}\, ,
\end{equation}
which are linked to Mandelstam invariants through $s_{ij}=\langle ij\rangle [ji]$ (see e.g.~\cite{Maitre:2007jq} for matching conventions). 
We will also use a shorthand for spinor chains,
\begin{equation}
  \langle i | j \pm k | i ] = \langle i j \rangle [ji] \pm \langle ik
    \rangle [ki] \, ,
\end{equation}
and we write $\trFive$ as\footnote{We note that $\trFive$ in ref.~\cite{Abreu:2021oya} differs by a minus sign compared to this definition.}
\begin{equation}
  \trFive = [12]\langle23\rangle[34]\langle41\rangle-\langle12\rangle[23]\langle34\rangle[41] \, .
\end{equation}

Under Lorentz transformations, spinor contractions have a residual
covariance under little-group transformations. In fact, a little-group
transformation of the $i^{th}$ leg with helicity $h_i$ reads
$(\lambda_i,\tilde\lambda_i)\rightarrow
(z_i \lambda_i,\tilde\lambda_i/z_i)$. Accordingly, helicity amplitudes
transform as $A \rightarrow z_i^{-2h_i} A$. We refer to the exponent
of the $z_i$ as the little-group weight. In summary,
helicity amplitudes are homogeneous expressions of spinor brackets not
just w.r.t.~the total degree but also w.r.t~each little-group weight.

Finally, throughout the article we will denote the group of cyclic
permutations of $n$ elements $\{i_1,\ldots{},i_n\}$ by ${{\cal Z}_n}( i_1,\ldots{},i_n ) $ and the set of all
permutations of $n$ elements, the symmetric permutation group, by
${\cal S}_n (i_1,\ldots{},i_n) $.

\subsection{Color space}\label{sec:amplitudes}

The external gluons are in the adjoint representation of $SU(N_c)$,
with indices denoted as $a,b,c$ or $a_k$, which run over over
$N_c^2-1$ values.  Quarks carry (anti-)fundamental color indices $i,
j$ ($\bar i, \bar j$) or $i_k$ ($\bar i_k$), which assume $N_c$
values.  Both color indices will be collected in tuples $\vec{a}
= \{a_1,\ldots{},a_n,i_1,\bar i_1 ,\ldots{} ,\bar i_n \}$.  We
explicitly represent the parton amplitudes in the color space
through the trace basis as in ref.~\cite{Bern:1990ux}.

We use the shorthand notation $\tr(i_1,\ldots{},i_n) = \tr(T^{a_{i_1}}\cdots{}T^{a_{i_n}})$,
and $(T^{a})^{\,\bar i}_i$ are the hermitian and traceless generators of
the fundamental representation of $SU(N_c)$.

The generators $T^a$ are normalized as,
\begin{equation}
{\tr}(T^a T^b) = \delta^{ab}\,,
\end{equation}
and fulfill the commutator relations,
\begin{align}
  \left[T^a,T^b\right] & =i f_{abc} T^c \, , \\
  i f_{abc} & =\tr(T^a T^b T^c) - \tr(T^b T^a T^c) \,.
\end{align}
The color algebra used in this paper is obtained from applying the  Fierz identity,
\begin{equation}
(T^a)^{\,\,\bar i}_i (T^a)^{\,\,\bar j}_j = 
\delta^{\bar i}_j \delta^{\bar j}_i - 
\frac{1}{N_c} \delta^{\bar i}_i \delta^{\bar j}_j\,.
\end{equation}
which allows to evaluate the summation over adjoint indices.

Using the above notation, the three-gluon two-quark scattering amplitudes are 
decomposed in color structures as
\begin{multline}\label{eq:partial_2q}
  \mathcal{A}_{\vec{a}}(1_u,2_{\bar u},3_g,4_g,5_g) = \\
 \sum_{\sigma \in \mathcal{S}_3(3,4,5)} \sigma\Big(
(T^{a_3}T^{a_4}T^{a_5})^{\,\bar i_2}_{i_1} \; 
A_{1}(1,2,3,4,5)\Big) \; + \\
 \sum_{\sigma \in \frac{\mathcal{S}_3(3,4,5)}{\mathcal{Z}_2(3,4)}} 
\sigma\Big(\tr(3,4) (T^{a_5})^{\,\bar i_2}_{i_1} 
\; A_{2}(1,2,3,4,5)\Big) \; + \\
 \sum_{\sigma \in \frac{\mathcal{S}_3(3,4,5)}{\mathcal{Z}_{3}(3,4,5)}} 
\sigma\Big(\tr(3,4,5) \delta^{\bar i_2}_{i_1} 
A_{3}(1,2,3,4,5)\Big) \; .
\end{multline}
Here $\sigma = \{\sigma_1,\ldots{},\sigma_5\}$ denotes permutations which acts on all
external-particle labels as $\sigma(i) = \sigma_i$. 
The sums run over all permutations that do not leave
the respective color structures invariant, i.e.\ over 6,
3, and 2 elements in the three lines of 
\cref{eq:partial_2q}, respectively.

Similarly, the one-gluon four-quark amplitudes are given by,
\begin{multline}\label{eq:partial_4q}
  \mathcal{A}_{\vec{a}}(1_u,2_{\bar u},3_d,4_{\bar d},5_g)   = \\
 \sum_{\sigma \in \mathcal{Z}_2(\{1,2\},\{3,4\})} \sigma\Big(
\delta^{\bar i_4}_{i_1} (T^{a_5})^{\,\bar i_2}_{i_3} 
\; A_{4}(1,2,3,4,5)\Big) \; + \\
 \kern-10mm \sum_{\sigma \in \mathcal{Z}_2(\{1,2\},\{3,4\})} \kern-2mm \sigma\Big(
\delta^{\bar i_2}_{i_1} (T^{a_5})^{\,\bar i_4}_{i_3} 
\; A_{5}(1,2,3,4,5)\Big)\,,\kern-1mm
\end{multline}
where the sums run over exchanging quarks pairs.

The amplitudes $A_i$ admit an expansion in terms of the bare QCD
coupling constant $\alpha_s^0 = (g_s^0)^2/(4\pi)$,
\begin{equation}\label{eq:as-series}
  A_i = (g_s^0)^3 \left(\sum_{L=0}^2 \left(\frac{\alpha_s^0}{2 \pi}\right)^L A_i^{(L)} 
	~+~ \order{(\alpha_s^{0})^3}\right) 
\end{equation}
with $L$ denoting the number of loops. 

\begin{widetext}
The amplitudes can be further
expanded in powers of $N_c$ and $N_f$ through two loops. 
For the two-quark amplitudes we obtain the following decomposition, 
\begin{subequations}\label{eq:Nc-Nf-series-2q}
\begin{align}
  A_1^{(0)} &= A_1^{(0),(0,0)} \,,  \quad A_2^{(0)} = 0\,, \quad A_3^{(0)} = 0\,, \\
  A_1^{(1)} &= N_c~A_1^{(1),(1,0)} + \frac{1}{N_c} A_1^{(1),(-1,0)} +N_f A_1^{(1),(0,1)}\,, 
	\quad A_2^{(1)} = A_2^{(1),(0,0)} +\frac{N_f}{N_c}A_2^{(1),(-1,1)}\,, \\ 
  A_3^{(1)} &= A_3^{(1),(0,0)} +\frac{N_f}{N_c} A_3^{(1),(-1,1)}\,, \\[1em]
  A_1^{(2)} &= N_c^2~A_1^{(2),(2,0)}+\red{A_1^{(2),(0,0)}}+\frac{1}{N_c^2}\red{A_1^{(2),(-2,0)}}
         + N_f N_c~A_1^{(2),(1,1)} + \frac{N_f}{N_c} \red{A_1^{(2),(-1,1)}} + N_f^2~A_1^{(2),(0,2)}\,, \\ 
  A_2^{(2)} &= N_c~\red{A_2^{(2),(1,0)}} + \frac{1}{N_c} \red{A_2^{(2),(-1,0)}} 
   	+  N_f\red{A_2^{(2),(0,1)}} + \frac{N_f}{N_c^2} \red{A_2^{(2),(-2,1)}} + \frac{N_f^2}{N_c} \red{A_2^{(2),(-1,2)}} \\
  A_3^{(2)} &= N_c \red{A_3^{(2),(1,0)}} + \frac{1}{N_c}\red{A_3^{(2),(-1,0)}} + N_f N_c\red{A_3^{(2),(1,1)}} 
	+ \frac{N_f}{N_c^2}\red{A_3^{(2),(-2,1)}} + \frac{N_f^2}{N_c}\red{A_3^{(2),(-1,2)}} \,.
\end{align}
\end{subequations}
Here, we remark that the helicity amplitudes $A_2^{(1),(-1,1)}$ and $A_2^{(2),(-1,2)}$ vanish.
The one-loop decomposition matches the one given in 
refs.~\cite{Bern:1994fz,DelDuca:1999rs} up to sign conventions.

For the four-quark amplitudes we find the decomposition, 
\begin{subequations}\label{eq:Nc-Nf-series-4q}
\begin{align}
  A_4^{(0)} &= A_4^{(0),(0,0)} \,, \quad A_5^{(0)} = \frac{1}{N_c} A_5^{(0),(-1,0)} \,, \\
  A_4^{(1)} &= N_c A_4^{(1),(1,0)} + \frac{1}{N_c} A_4^{(1),(-1,0)} + N_f A_4^{(1),(0,1)} \,,\quad \\
  A_5^{(1)} &= A_5^{(1),(0,0)} + \frac{1}{N_c^2} A_5^{(1),(-2,0)} + \frac{N_f}{N_c} A_5^{(1),(-1,1)} \,, \\[1em]
  A_4^{(2)} &= N_c^2 A_4^{(2),(2,0)} + \red{A_4^{(2),(0,0)}} + \frac{1}{N_c^2} \red{A_4^{(2),(-2,0)}}
   	 +  N_f N_c A_4^{(2),(1,1)} + \frac{N_f}{N_c} \red{A_4^{(2),(-1,1)}} + N_f^2  A_4^{(2),(0,2)} \\
  A_5^{(2)} &= N_c\red{A_5^{(2),(1,0)}}+\frac{1}{N_c}\red{A_5^{(2),(-1,0)}}+\frac{1}{N_c^3}\red{A_5^{(2),(-3,0)}}
   	 + N_f\red{A_5^{(2),(0,1)}} + \frac{N_f}{N_c^2} \red{A_5^{(2),(-2,1)}} + \frac{N_f^2}{N_c} \red{A_5^{(2),(-1,2)}}\,.
\end{align}
\end{subequations}
\end{widetext}

The coefficients $A_i^{(L),(n_c,n_f)}$ will be called \emph{partial amplitudes}.
In the limit of large number of colors, with the ratio $N_f/N_c$
fixed, only the partial amplitudes with $L=n_c+n_f$
contribute \cite{tHooft:1973alw}, receiving contributions only
from planar diagrams. These leading-color partial amplitudes have been
calculated in refs.~\cite{Badger:2018gip,Abreu:2018jgq,Abreu:2019odu}.
The remaining amplitudes, marked in red, receive contributions from
non-planar diagrams and are the new result of this work.  For
convenience we also recalculate all previously known amplitudes in
\cref{eq:Nc-Nf-series-2q,eq:Nc-Nf-series-4q}.

\subsection{Renormalization}

We employ the 't~Hooft--Veltman scheme of dimensional regularization to regularize ultraviolet (UV) and infrared (IR) divergences of loop amplitudes, where the
number of space-time dimensions is set to $D=4-2\epsilon$. For helicity amplitudes with external quarks we employ the prescription of ref.~\cite{Abreu:2018jgq}.
To cancel the UV divergences, we renormalize the bare QCD coupling constant in the 
$\MSbar$ scheme. This is accomplished by the replacement in \cref{eq:as-series},
\begin{align}\label{eq:renorm}
  &\alpha_0\mu_0^{2\epsilon}S_{\epsilon}
  =\alpha_s\mu^{2\epsilon} \times \\
  &\left(
  1-\frac{\beta_0}{2\epsilon}\frac{\alpha_s}{2\pi}
  +\left(\frac{\beta_0^2}{4 \epsilon^2}-\frac{\beta_1}{ 8 \epsilon}\right) \left(\frac{\alpha_s}{2\pi}\right)^2+\mathcal{O}\left(\alpha_s^3\right)\right)\,, \nonumber
\end{align}
where $S_\epsilon=(4\pi)^{\eps}e^{-\eps\gamma_E}$, with
$\gamma_E=-\Gamma'(1)$ the Euler-Mascheroni constant, 
and $\mu_0,\mu$ are regularization and renormalization scale respectively.
The QCD $\beta$-function's expansion coefficients are
\begin{subequations}
  \begin{align}
    \beta_0&=\frac{11}{3} N_c -  \frac{2}{3} N_f \;\, , \\
    \beta_1&=\frac{34}{3} N_c^2 - \frac{13}{3} N_c N_f + \frac{N_f}{N_c} \,.
  \end{align}
\end{subequations}
We then expand the renormalized amplitudes through the renormalized coupling as in \cref{eq:as-series}.

The remaining divergences are of IR origin and can be predicted by the universal factorization 
\cite{Catani:1998bh,Sterman:2002qn,Becher:2009cu,Gardi:2009qi}:
\begin{equation}\label{eqn:remainder}
  {\cal R}(\mu) ={\bf Z}(\epsilon, \mu) {\cal A}(\mu) ~+~ \order{\epsilon} \,,
\end{equation}
where the \emph{finite remainder} ${\cal R}$ is obtained by the application of
the color-space operator ${\bf Z}$, which is derived by exponentiation \cite{Becher:2009cu},
\begin{align}
{\bf Z}^{-1} (\epsilon, \mu)={\bf P}\, {\rm exp}\left[\int^\infty_{\mu}\frac{d\mu^\prime}{\mu^\prime} {\bf \Gamma}(\mu^\prime)\right],
\end{align}
of the anomalous dimension matrix
\begin{align} \nonumber
&\kern-5mm{\bf\Gamma}(\mu)= -\sum_{(i,j)} {\bf T}_i\cdot {\bf T}_j \times \\
&\kern5mm\frac{\gamma_{\rm cusp}}{2} \; {\rm ln}\left(-\frac{s_{ij}}{\mu^2} - \ii 0\right) ~+~ n_q \gamma^{q} + n_g \gamma^{g} .
\end{align}
Here the sum runs over pairs of external partons,
and the color-space operators $\mathbf{T}_i$ act on the color representation of the
i$^{th}$ parton. 
For adjoint indices the action is given by $({\bf T}_i^a)_{bc}=-i f^{abc}$,
and for fundamental indices as $({\bf T}^a_i)_j^{\,\bar k}= \pm (T^a)_j^{\,\bar k}$;
$n_q$ and $n_g$ are the number of quarks and gluons in the process respectively,
and the functions $\gamma_{\mathrm{cusp}},\gamma^q =\gamma^{\bar{q}},\gamma^g$ can 
be found in ref.~\cite[Appendix A]{Becher:2009qa}\footnote{
  The rescaling by a factor of 2 per loop is required to match our expansion in $\alpha_s/2\pi$ in \cref{eq:as-series}.
}.

After absorbing all divergences of amplitudes into UV and IR renormalization through
\cref{eq:renorm,eqn:remainder} we recover expansions of
\cref{eq:as-series,eq:partial_2q,eq:partial_4q,eq:Nc-Nf-series-2q,eq:Nc-Nf-series-4q}
at the level of finite remainders $\mathcal{R}$.
We then define \emph{partial finite remainders}
\begin{align}\label{eqn:finRemainder}
  R_{j, \vec{h}}^{(L),(n_c,n_f)}(i_1,\ldots{},i_5),
\end{align}
which will be the elementary building blocks considered in this work.

\subsection{Generating set of finite remainders}
\label{subsec:generating-partial-helicity-remainders}

Using parity, charge conjugation transformations and permuting momentum
assignments to the external states, we find a generating set of
finite remainders that we need to compute.
Identities inherited from color factors allow one to further restrict the set of required functions, as discussed below \cref{sec:colorIdentities}.

Focusing first on two-loop data and suppressing the labels specifying the $N_c$
and $N_f$ decomposition we consider a generating set of helicity assignments.
To this end we generate all helicity assignments, and select one
representative from the orbits of the charge conjugation and parity transformations.
Furthermore, given that we are considering analytic amplitudes, we chose a
convenient momentum assignment for each helicity amplitude,
which we line up with the little group transformation properties in the single-minus and the MHV amplitudes.

For the single-minus helicity configuration we have
\begin{subequations}\label{eqn:RsingleMinus}
  \begin{align}
  &R_1(1^+,2^-,3^+,4^+,5^+) \,,\\
  &R_2(1^+,2^-,3^+,4^+;5^+) \,,\\
  &R_3(1^+,2^-,3^+,4^+,5^+) \,.
  \end{align}
\end{subequations}
For the MHV configurations we have five generating remainders,
\begin{subequations}\label{eqn:Rmhv-2q}
  \begin{align}
  &R_1(1^+,2^-,3^+,4^+,5^-) \,,\\
  &R_1(1^+,2^-,3^+,5^-,4^+) \,,\\
  &R_1(1^+,2^-,5^-,4^+,3^+) \,,\\
  &R_2(1^+,2^-,3^+,4^+,5^-) \,,\\
  &R_2(1^+,2^-,3^+,5^-,4^+) \,,\\
  &R_3(1^+,2^-,3^+,4^+,5^-) \,.
  \end{align}
\end{subequations}
The analogous analysis yields the following generating set of finite four-quark remainders with
the MHV helicity configuration,
\begin{subequations}\label{eqn:Rmhv-4q}
  \begin{align}
  &R_4(1^+,2^-,3^+,4^-,5^+) \,,\\
  &R_4(1^+,2^-,4^-,3^+,5^+) \,,\\
  &R_4(2^-,1^+,3^+,4^-,5^+) \,,\\
  &R_5(1^+,2^-,3^+,4^-,5^+) \,,\\
  &R_5(1^-,2^-,4^-,3^+,5^+) \,.
  \end{align}
\end{subequations}
In \cref{sec:colorIdentities} we discuss further identities between remainders associated 
to distinct terms in the $N_c$,$N_f$ decomposition.

\subsection{NNLO hard function}

The two-loop NNLO QCD corrections for partonic cross sections 
are obtained from squared helicity- and color-summed partial remainders, which we call \emph{hard functions} $\mathcal{H}$,
\begin{equation}\label{eq:hard-function}
  \mathcal{H}= \frac{1}{\mathcal{B}} \sum_{\vec h, \vec a} \abs{{\cal R}_{\vec h,\vec a}}^2 \,, \quad \mathcal{B} =  \sum_{\vec h, \vec a} \abs{\mathcal{A}^{(0)}_{\vec{h},\vec{a}}}^2 \,.
\end{equation}
Here the summation is performed by mapping each partial remainder into one from
the generating sets \cref{eqn:RsingleMinus,eqn:Rmhv-2q,eqn:Rmhv-4q}.

The hard function is expanded perturbatively up to $\order{\alpha_s^2}$ as in \cref{eq:as-series}. 
We further expand $\mathcal{H}$ in powers of $N_f$, while we keep the value of $N_c$ implicit,\footnote{
  In ref.~\cite{Abreu:2021oya} a factor of $\frac{N_c}{2}$ is additionally extracted at each order.
}
\begin{subequations}
  \label{eq:hard-function-partial}
  \begin{align}
    \mathcal{H}^{(0)} &= 1  \,,  \\
    \mathcal{H}^{(1)} &= H^{(1)[0]} ~+~ N_f H^{(1)[1]} \,, \\
    \mathcal{H}^{(2)} &= H^{(2)[0]} ~+~ N_f H^{(2)[1]}~+~ N_f^2 H^{(2)[2]} \,.
  \end{align}
\end{subequations}

\subsection{IR scheme change}
\label{sec:scheme-change}

It is important to highlight that the finite remainders encompass all the physical details related to the underlying scattering process.
Specifically, one can compute any observable by utilizing finite remainders (see e.g.~\cite{Weinzierl:2011uz}).
In this way it is possible to cancel many undesirable side effects of dimensional regularization.

In this work we use the \emph{minimal subtractions} scheme of IR
renormalization, following the conventions of
refs.~\cite{Becher:2009cu,Becher:2009qa}.  One might be interested in obtaining
finite remainders defined in a different IR renormalization scheme, e.g.~
Catani's scheme \cite{Catani:1998bh,Gardi:2009qi,Becher:2009cu,Becher:2009qa}.
In the following we show that it is possible to convert the finite remainders
in our scheme to any other scheme by an additional \emph{finite
renormalization}, i.e.\ the knowledge of higher orders in $\epsilon$ of
amplitudes is only initially required to derive finite remainders in arbitrary
scheme.  We take advantage of this fact in our computational framework and
circumvent analytic reconstruction of amplitudes.

Suppose we are interested in a different scheme where the finite remainder is
defined as 
\begin{equation}\label{eqn:remainder2} 
\tilde{\mathcal{R}} =
\tilde{\mathbf{Z}}(\epsilon) {\cal A}(\epsilon) ~+~ \order{\epsilon} \,,
\end{equation} 
where ${\cal A}$ is the same UV renormalized amplitude as in
\cref{eqn:remainder}.  Here we remind the reader that $\mathcal{A}$ and
$\mathcal{R}$ are vectors and $\tilde{\mathbf{Z}}$ is an operator in color
space.  We assume that both $\mathbf{Z}$  and $\tilde{\mathbf{Z}}$ have a
perturbative expansion as in \cref{eq:as-series}, and
$\mathbf{Z},\tilde{\mathbf{Z}} = 1 + \order{\alpha_s} $.

We then consider the difference 
\begin{equation}\label{eq:ir-scheme-change}
  \delta\mathcal{R} = \mathcal{R} - \tilde{\mathcal{R}} 
	= \left(\mathbf{Z} - \tilde{\mathbf{Z}}\right)\mathbf{Z}^{-1}  ~ \mathcal{R}  
	= \delta\mathbf{Z} ~ \mathcal{R},
\end{equation}
which is finite by definition.  Therefore the operator $\delta\mathbf{Z} =
\left(\mathbf{Z} - \tilde{\mathbf{Z}}\right)\mathbf{Z}^{-1}$ must not have
$\epsilon$ poles, except possibly the ones that cancel upon action on
$\mathcal{R}$.  Provided the latter cancellation does not rely on the existence
of a non-trivial null space, both $\delta\mathbf{Z}$ and $\mathcal{R}$ and can
be truncated at $\order{\epsilon^0}$.  We can therefore express the remainder
$\tilde{\mathcal{R}}$ through $\mathcal{R}$ order-by-order in $\alpha_s$ by a
finite renormalization $\delta\mathbf{Z}$, whose perturbative expansion starts
at $\order{\alpha_s}$. 

For the squared finite remainders we can write more explicitly through two loops 
\begin{subequations}
  \begin{align}
    & \abs{\tilde{\mathcal{R}}_{\vec{h}, \vec{a}}^{(0)}}^2  = \abs{\mathcal{R}_{\vec{h}, \vec{a}}^{(0)}}^2 \,, \\ 
    & 2 \Re\Big[\mathcal{R}^{(0)\dagger}_{\vec{h}, \vec{a}}\tilde{\mathcal{R}}_{\vec{h}, \vec{a}}^{(1)}\Big]  = 2 \Re\Big[\mathcal{R}^{(0)\dagger}_{\vec{h}, \vec{a}}\mathcal{R}_{\vec{h}, \vec{a}}^{(1)}\Big]  \,-\\ \nonumber
    & \qquad 2  \Re\Big[ \mathcal{R}^{(0)\dagger}_{\vec{h}, \vec{a}} \delta\mathbf{Z} ^{(1)}\mathcal{R}^{(0)}_{\vec{h}, \vec{a}} \Big] \,, \\ 
    &  \abs{\tilde{\mathcal{R}}_{\vec{h}, \vec{a}}^{(1)}}^2 = \abs{\mathcal{R}_{\vec{h}, \vec{a}}^{(1)}}^2 \,+ \\\nonumber
    & \quad   \mathcal{R}^{(0)\dagger}_{\vec{h}, \vec{a}} \delta\mathbf{Z} ^{(1)\dagger}\delta\mathbf{Z} ^{(1)}\mathcal{R}^{(0)}_{\vec{h}, \vec{a}} 
      -  2  \Re\Big[ \mathcal{R}^{(0)\dagger}_{\vec{h}, \vec{a}} \delta\mathbf{Z} ^{(1)}\mathcal{R}^{(1)}_{\vec{h}, \vec{a}} \Big],\\ 
    & 2 \Re\Big[\mathcal{R}^{(0)\dagger}_{\vec{h}, \vec{a}}\tilde{\mathcal{R}}_{\vec{h}, \vec{a}}^{(2)}\Big]  = 2 \Re\Big[\mathcal{R}^{(0)\dagger}_{\vec{h}, \vec{a}}\mathcal{R}_{\vec{h}, \vec{a}}^{(2)}\Big]  \,-\\\nonumber 
    & \qquad  2  \Re\Big[ \mathcal{R}^{(0)\dagger}_{\vec{h}, \vec{a}} \delta\mathbf{Z} ^{(2)}\mathcal{R}^{(0)}_{\vec{h}, \vec{a}} \Big] \,.
  \end{align}
\end{subequations}
It is then straightforward to perform helicity and color summation to derive
hard functions in the new scheme.

We have explicitly calculated the operator $\delta\mathbf{Z}$ to convert the
minimal subtractions scheme to the Catani scheme.  We verified that the latter
must be supplemented by both types of tripole color correlation terms added in
the later revisions of \cite[eq.~(17)]{Becher:2009qa} for the poles to cancel
at the level of partial remainders for five-parton scattering. 
In addition, we have verified numerically that the $\mathcal{O}(\alpha_s^2)$ contributions to the squared helicity- and color-summed matrix elements
originating from the tripole correlation terms vanish in all five-parton channels.

\section{Numerical sampling of remainders}
\label{sec:numSamples}

We will construct the partial remainders from analytic reconstruction, i.e. we
compute the analytic form of the partial remainders from numerical evaluations
in a finite field.  Partial remainders can be expressed as a linear combination
of transcendental integral functions $h_i$ and rational coefficient functions
$r_i$,
\begin{align}\label{eq:funcBasis}
R=\sum_i r_i h_i \,.
\end{align}
The integral functions, referred to as pentagon function, are known
\cite{Chicherin:2020oor}. The computation of the analytic form of the function
coefficients $r_i$ is one of the central results of this paper.

Here we summarize the input data required for our computation following
ref.~\cite{DeLaurentis:2023nss}.
For numerical evaluation of remainder functions in a finite field we use the
program \textsc{Caravel} \cite{Abreu:2020xvt}, which implements the multi-loop
numerical unitarity method \cite{Ita:2015tya,Abreu:2017xsl,Abreu:2017hqn}.  In
this approach amplitudes are reduced to a set of master integrals by matching
numerical evaluations of generalized unitarity cuts to a parametrization of the
loop integrands. For the five-parton process we use the recently obtained
non-planar parametrization \cite{Abreu:2023bdp,DeLaurentis:2023nss}.  
Furthermore, for the quark processes
we extended the set of planar unitarity cuts to non-planar diagrams which are
required for subleading-color partial amplitudes.  We generated the cut
diagrams with \texttt{qgraf} \cite{Nogueira:1991ex} and arranged them into a
hierarchy of cuts with a private code. 
The unitarity cuts evaluated through color-ordered tree amplitudes are matched
to the amplitude definitions in \cref{eq:partial_2q,eq:partial_4q}, by
employing the unitarity based color decomposition
\cite{Ochirov:2016ewn,Ochirov:2019mtf}.  The $\epsilon$-dependence of cuts that
originates from the state sums in loops are obtained by the 
dimensional reduction method developed in ref.~\cite{Anger:2018ove,Abreu:2019odu,Sotnikov:2019onv}.
With these upgrades \textsc{Caravel} now computes the function coefficients
$r_i$ of five-parton partial amplitudes up to two loops, given a kinematic
point and a choice of polarization labels for the external gluons. 

Next we will require two types of numerical samples of the 
remainder functions which we repeat from ref.~\cite{DeLaurentis:2023nss} for convenience:
\begin{enumerate}
\item Random phase-space points: these are $N$ randomly generated phase-space
points which we label by the superscript $n$. We represent these points in terms of 
sets of spinor variables,
\begin{align}\label{eqn:randomPS}
\{ \{\lambda_1^{n}, ... , \lambda_5^{n},\tilde \lambda_1^{n}, ... , \tilde \lambda_5^{n} \} \}_{n=1,N} \,.
\end{align}
They are subject to momentum conservation $\sum_i \lambda_i^n\tilde\lambda^n_i=0$.
Below we will use the shorthand notation,
\begin{align}\label{eqn:randomPSVec}
 \vec \lambda = \{\lambda_1, ... , \lambda_5,\tilde \lambda_1, ... , \tilde \lambda_5 \} \,.
\end{align}
to denote the spinor-helicity variables associated to a phase space point.

\item (Anti-)holomorphic slice:
this is one holomorphic slice
\cite{PageSAGEXLectures,Abreu:2021asb,Abreu:2023bdp}
associated to a random phase-space point (\ref{eqn:randomPS}),
\begin{align}
\label{eqn:famHolomorphicSlice}
&\lambda_i(t)=\lambda_i+t c_i\eta\,,\quad 
\tilde\lambda_i(t) = \tilde\lambda_i\,,\\ \nonumber
&\qquad \sum_{i=1}^5 c_i \tilde\lambda_i=0\,. 
\end{align}
Here the reference spinor $\eta$ is chosen randomly. The constants $c_i$ are 
obtained by solving the linear momentum-conservation condition. 
In addition we will use the anti-holomorphic slice which is obtained from 
\cref{eqn:famHolomorphicSlice} by swapping $\lambda_i\leftrightarrow
\tilde\lambda_i$, $\eta \rightarrow \tilde\eta$ 
and renaming $t\rightarrow \bar t$.
\end{enumerate}

\section{Identities between partial amplitudes}
\label{sec:colorIdentities}

Partial amplitudes in the trace-basis representation are known 
to satisfy linear relations originating from symmetry properties 
and the adjoint representation gauge interactions of field theory.
These relations, which we will refer to as \emph{color identities}, 
can be exploited to reduce the number of partial 
amplitudes that need to be computed, and subsequently to improve 
the efficiency of the numerical evaluation 
of the hard functions (\ref{eq:hard-function}).

Color identities were discussed for multi-loop gluon
amplitudes \cite{Edison:2011ta} and recently in ref.~\cite{Dunbar:2023ayw}. 
Much less is known
about the scattering of quarks and gluons at two loops. Here we empirically
identify all linear relations for the five-point amplitudes including quarks
from numerical evaluations. We proceed as follows.

Linear relations can only hold between remainders with the same little-group weight, which we specify below by the labels $\vec h=\{h_1,...,h_5\}$.
We group all remainders with identical little-group weights into sets
\begin{align}
P^{(L),\vec h} = 
\big\{ R_{k_i}
^{(L),(n_c^i,n_f^i)}(\vec h, \vec \lambda) \big\}_{i=1,S} \,,
\end{align}
where the index $i$ enumerates the $S$ remainder functions. The superscript $\vec
\lambda$ specifies the momentum of the states (see \cref{eqn:randomPS}).

We employ random numerical samples (\ref{eqn:randomPS}),
which associate a vector of numerical values to each remainder,
\begin{align}
\big\{ 
	R_{k_i}^{(L),(n_c^i,n_f^i)}(\vec h, \vec \lambda^n) 
\big\}_{i=1,S ;n=1,N}\,.
\end{align}
We then search for vanishing linear combinations of these vectors,
\begin{align}
\sum_{i=1}^S 
	R_{k_i}^{(L),(n_c^i,n_f^i)}(\vec h, \vec \lambda^n) \, c_i=0 \,, \quad\mbox{for}\quad  n=1, N \,.
\end{align}
The set of nontrivial constant solutions $c_i$ yield 
the desired identities between partial remainders. 

Technically, we simplify the search for identities by using finite-field
arithmetic and exploiting the fact that the transcendental functions in the decomposition
(\ref{eq:funcBasis}) form a basis \cite{Chicherin:2021dyp}.
We first obtain identities between coefficients of selected transcendental functions. 
We then intersect them in order to find an identity valid for the
entire remainder.

In principle we can  distinguish two classes of identities: 1) helicity dependent ones, which
hold in a single class of helicity configurations, i.e.~in the single-minus or
MHV configuration. 2) Helicity independent relations, which hold for all
helicity assignments. Color identities are of the second type. 
At two loop we find only helicity independent identities.

Let us note that algorithms to obtain color decompositions and relations are
well understood for arbitrary multiplicity at one loop
\cite{DelDuca:1999rs,Reuschle:2013qna,Ita:2011ar,Badger:2012pg,Kalin:2017oqr}.
In particular, representations of two-quark and four-quark amplitudes in terms
of so called primitive amplitudes are known
\cite{Bern:1994fz,DelDuca:1999rs,Kunszt:1994nq,Ellis:2008qc}, which imply the
identities that we study here.

\subsection{Two-quark channel}
\label{subsec:2q3gIdentities}

We consider first the remainders associated to the partial amplitudes of ${\cal
A}(1^{h_1}_u, 2^{h_2}_{\bar u}, 3^{h_3}_g, 4^{h_4}_g, 5^{h_5}_g)$
\cref{eq:partial_2q}. To start with, we collect the set of all remainders with
identical little-group weights.  These are obtained from evaluating the remainders
on permuted momenta, keeping helicity quantum numbers assigned to each momentum
fixed. 
With this in mind, the full group of permutations is generated by the groups
$\mathcal{S}_2(1^{h_1}, 2^{h_2})$ and $\mathcal{S}_3(3^{h_3},4^{h_4},5^{h_5})$,
which do not mix gluons and quarks. In total, the permutation group contains
$2\times6=12$ elements.

This set of remainders can be further reduced, using symmetry properties 
of the color factors:
\begin{enumerate}
\item[$R_1$:] Charge conjugation symmetry considered for the amplitude ${\cal
A}(1^{h_1}_u, 2^{h_2}_{\bar u}, 3^{h_3}_g, 4^{h_4}_g, 5^{h_5}_g)$ 
(\ref{eq:partial_2q}) forces the
coefficients of the color factors $(T^{a_3,a_4,a_5})_{i_1}^{\,\bar i_2}$ and
$(T^{a_5,a_4,a_3})_{i_2}^{\,\bar i_1}$ to match, e.g.
$R_1(1,2,3,4,5)=-R_1(2,1,5,4,3)$. This relation halves the number of independent
momentum permutations. We chose representatives generated by
$\mathcal{S}_3(3^{h_3},4^{h_4},5^{h_5})$, namely
\begin{align} 
& R_1(1,2,3,4,5),\, R_1(1,2,3,5,4), \nonumber \\
& R_1(1,2,4,3,5),\, R_1(1,2,4,5,3), \nonumber \\
& R_1(1,2,5,3,4),\, R_1(1,2,5,4,3) \, .  
\end{align}
\item[$R_2$:] Using charge conjugation and the cyclicity of the trace
$\tr(3,4)=\tr(4,3)$ implies that partial amplitudes $A_2$ (and their
remainders) are unchanged under $S_2(3^{h_3},4^{h_4})$, and
$\mathcal{S}_2(1^{h_1}, 2^{h_2})$.  Its symmetry group thus has dimension 4,
meaning that out of the total 12 permutations of momenta (and helicities) there
are 3 independent permutations.  We chose the representatives 
\begin{align} &
R_2(1,2,3,4,5),\, R_2(1,2,3,5,4), \nonumber \\ & R_2(1,2,4,5,3) \, .
\end{align}

\item[$R_3$:]
Finally, charge conjugation symmetry and cyclicity of the trace
$\tr(3,4,5)$ implies invariance of the partial amplitudes $A_3$ (and $R_3$)
under the transformations $\mathcal{Z}_3(3^{h_3},4^{h_4},5^{h_5})$ and 
$\mathcal{S}_2(1^{h_1}, 2^{h_2})$.
After modding the 12 total permutations by these $3\times 2=6$ symmetry
transformations 2 independent permutations remain, which we pick to be 
\begin{equation} R_3(1,2,3,4,5),\, R_3(1,2,3,5,4) \, .  \end{equation} 
\end{enumerate}
Here we suppressed again the superscripts specifying the partial remainder, 
namely their loop order and contribution in $N_c$ and $N_f$ in
\cref{eq:partial_2q}.

We have obtained a set of two-loop partial remainders with identical little-group weight,
\begin{align}\label{eq:2qSet}
P^{(2),\vec h}=\{& R_1^{(2),(2,0)}(1^{h_1},2^{h_2},3^{h_3},4^{h_4},5^{h_5}) , \ldots ,\nonumber \\ 
	&\quad  R_3^{(2),(-1,2)}(1^{h_1},2^{h_2},3^{h_3},5^{h_5},4^{h_4}) \} \,,
\end{align}
All remainders in this set of remainders can be expressed in 
terms of the generating set of \cref{subsec:generating-partial-helicity-remainders}.
After following the steps discussed in the beginning of this section we find no nontrivial 
relations in the two-loop two-quark channel.

\subsection{Four-quark channel}
\label{subsec:4q1gIdentities}

The full group of permutations that maintains the little-group weight of 
$A(1_u, 2_{\bar u}, 3_d, 4_{\bar d},
5_g)$ is generated by the following cycles ${\cal S}_2(1^{h_1},2^{h_2})$, 
${\cal S}_2(3^{h_3}, 4^{h_4})$ and the transformation $(\{1_u, 2_{\bar
u}\} \leftrightarrow \{3_d, 4_{\bar d}\})$. In total, the permutation
group contains 8 elements.

Charge conjugation implies the relations
$R_4(1,2,3,4,5)=-R_4(4,3,2,1,5)$ and $R_5(1,2,3,4,5)=-R_5(2,1,4,3,5)$,
as seen from hermitian conjugation of the color matrices in
\cref{eq:partial_4q}.
Hence, modding out the full permutation group of dimension 8, by the
group which leaves the partial remainders unchanged, up to a sign, we obtain the
following inequivalent set of partial remainders, 
suppressing the labels $L\,,n_c$ and $n_f$,
\begin{equation}
\label{eq:mixed-quark-line-color-permutations} 
\begin{gathered}
 R_4(1,2,3,4,5),\, R_4(1,2,4,3,5),\, \\
 R_4(2,1,3,4,5),\, R_4(2,1,4,3,5) \, , 
\end{gathered}
\end{equation}
and,
\begin{equation}
\label{eq:same-quark-line-color-permutations}
\begin{gathered}
R_5(1,2,3,4,5),\, R_5(1,2,4,3,5),\\
 R_5(3,4,2,1,5),\, R_5(4,3,2,1,5) \, . 
\end{gathered}
\end{equation}
The set of partial remainders among which linear relations may be found is then,
\begin{align}\label{eq:4qSet}
P^{(2)}=\{ &R_4^{(2),(2,0)}(1,2,3,4,5) , \ldots, \nonumber \\ 
	& \quad R_5^{(2),(-1,2)}(4,3,2,1,5) \}\,.
\end{align}

In contrast to the two-quark channel, we find nontrivial identities which to 
the best of our knowledge have not been reported previously:
\begin{widetext}
\begin{gather}
\label{eq:4q1g-2l-identity-nf0-singlet}
\begin{aligned}
&\kern-30mm \Big\{ \big[ 16 \, R^{(2),(2,0)}_4\, (1,2,3,4,5) 
+ 4 \, R^{(2),(0,0)}_4\, (1,2,3,4,5) + \\ \kern+30mm
& R^{(2),(-2,0)}_4(1,2,3,4,5) \big]
- \big[\dots \big]_{3 \leftrightarrow 4} \Big\}
- \Big\{ \dots \Big\}_{1 \leftrightarrow 2} = 0 \, .
\end{aligned} \\[2em] 
\label{eq:4q1g-2l-identity-nf0-doublet}
\begin{aligned}
&\kern-10mm \big[  32 \, R^{(2),(2,0)}_4\, (1,2,3,4,5) + 8 \, R^{(2),(0,0)}_4\, (1,2,3,4,5)
+ 2 R^{(2),(-2,0)}_4(1,2,3,4,5) \\ \kern+10mm
& + 16 \, R^{(2),(1,0)}_5\, (1,2,3,4,5) \, + 4 R^{(2),(-1,0)}_5(1,2,3,4,5) +   R^{(2),(-3,0)}_5 (1,2,3,4,5) \big]
  - \big[ \dots \big]_{3 \leftrightarrow 4}=  0 \, .
\end{aligned} \\[2em] 
\label{eq:4q1g-2l-identity-nf1-singlet}
\begin{aligned}
\Big\{ \big[ 4 R^{(2),(1,1)}_4(1,2,3,4,5)
+ R^{(2),(-1,1)}_4(1,2,3,4,5) \big] - \big[\dots \big]_{3 \leftrightarrow 4} \Big\} - \Big\{ \dots \Big\}_{1 \leftrightarrow 2}  = 0 \, .
\end{aligned} \\[2em]
\label{eq:4q1g-2l-identity-nf1-doublet}
\begin{aligned}
&\kern-30mm \big[  8 R^{(2),(1,1)}_4(1,2,3,4,5)) + 2 R^{(2),(-1,1)}_4(1,2,3,4,5) \\ \kern+30mm
& + 4 R^{(2),(0,1)}_5(1,2,3,4,5) \, + R^{(2),(-2,1)}_5(1,2,3,4,5) \big]
  - \big[ \dots \big]_{3 \leftrightarrow 4} =  0 \, .
\end{aligned}
\end{gather}
\end{widetext}

The identities (\ref{eq:4q1g-2l-identity-nf0-singlet}) and
(\ref{eq:4q1g-2l-identity-nf1-singlet}) are singlets under the
permutation group of \cref{eq:mixed-quark-line-color-permutations},
while the identities (\ref{eq:4q1g-2l-identity-nf0-doublet}) and
(\ref{eq:4q1g-2l-identity-nf1-doublet}) are doublets under the
permutation group of \cref{eq:same-quark-line-color-permutations}. The
former are manifestly anti-symmetric under ${\cal S}_2(1,2)$ and
${\cal S}_2(3,4)$, while the latter are manifestly anti-symmetric
under ${\cal S}_2(3,4)$.

As a final point, we note that the identities that we have found do
not allow us to express any of the partial remainders in terms of sums
over permutations of the others, i.e.~the generating set discussed
in \cref{subsec:generating-partial-helicity-remainders} cannot be
further reduced for any $n_c,\,n_f$.  This is in contrast to the
well-known fact that in the five-gluon channel the most subleading in
$N_c$ expansion partial amplitude can be
eliminated \cite{Edison:2011ta}.

\section{Analytic reconstruction}
\label{sec:reconstruction}

As discussed in section \ref{sec:numSamples}
we have available numerical evaluations of the
coefficients $r_i$ in the remainder function (\ref{eq:coeffBasis}). 
We will now build upon ref.~\cite{DeLaurentis:2023nss} to obtain compact
analytic expressions for the function coefficients from such numerical samples.
The starting point is the understanding that the coefficients admit the 
least common denominator representation, 
\begin{align}\label{eqn:LCDimproved}
r_i = \frac{{\cal N}_i(\lambda,\tilde \lambda)}{\prod_{j}
{\cal D}_j^{q_{ij}}(\lambda,\tilde \lambda)}\,,
\end{align}
where the denominator factors ${\cal D}_j$ are given by the letters in the symbol 
alphabet of pentagon functions
\cite{Chicherin:2017dob,Chicherin:2018old,Abreu:2018aqd} with integer exponents
$q_{ij}$ \cite{Abreu:2018zmy}. 
The goal of analytic reconstruction is to determine $q_{ij}$ and the
numerator polynomials ${\cal N}_i$ in \cref{eqn:LCDimproved}.

First we determine the exponents $q_{ij}$.
To this end we follow the 
univariate-slice reconstruction \cite{Abreu:2018zmy} in spinor-helicity
variables \cite{PageSAGEXLectures,Abreu:2021asb,Abreu:2023bdp}.  In this
approach the function coefficients are obtained as univariate rational functions 
$r_i(t)$ and $r_i(\bar t\,)$ on a holomorphic and an anti-holomorphic 
slice (\ref{eqn:famHolomorphicSlice}), respectively. Given the rational
functions $r_i(t)$ and $r_i(\bar t\,)$, their denominators are matched to
products of the letter polynomials ${\cal D}_j(t)$ and ${\cal D}_j(\bar t\,)$.  
This uniquely fixes the exponents $q_{ij}$ in each of the
functions $r_i$ (\ref{eqn:LCDimproved}). In particular considering holomorphic
and anti-holomorphic slices independently, ensures that one identifies purely
(anti-)holomorphic terms, such as $[ij]$ and $\langle i\, j \rangle$. 

The importance of the exponents $q_{ij}$ is two-fold.  On the one hand, we
determine which of the letters actually appears as denominator factors.  
For the five-parton finite remainders we observe that the set of
denominator factors consists of the $35$ elements,
\begin{align}\label{eq:codimension-one-poles}
  {\cal D} =& 
  \big\{\langle ij \rangle, 
	\, [ij], \,
	\langle i|j+k|l] ,... \big\}
\end{align}
where the set runs over all independent permutations of the spinor
strings/brackets. None of the coefficients in the remainder has a $\text{tr}_5$
singularity. 
On the other hand, the exponents constrain the ansatz (\ref{eqn:LCDimproved}),
since the mass dimension and little-group weight of the numerator polynomial ${\cal
N}_i$ is uniquely fixed by those of the helicity
amplitude and the denominator. Consequently, we can construct a finite
dimensional ansatz for the polynomial ${\cal N}_i$. We have thus reduced the
computation to the problem of finding finitely many polynomial parameters. 

Before we obtain the functions $r_i$ we wish to identify linear dependences, to
identify a minimal set of functions that we need to compute. To this end we
first sort the functions $r_i$ according to complexity, namely, the mass
dimension of the respective numerators ${\cal N}_i$, which in turn is
correlated with the polynomial's parameters. 
Next we identify linear dependence numerically via Gaussian elimination, and determine the indices of the basis coefficient functions in the set $B$, 
\begin{align}\label{eqn:funcBasis}
\{ r_i \}_{i\in B}\,.
\end{align}
In this way we further reduce the data,
required to specify the scattering process to a basis of rational coefficient
functions $r_i$ and a constant rational-valued matrix $M_{ij}$ \cite{Abreu:2019odu},
\begin{align}\label{eq:coeffBasis}
R=\sum_{j\in B,i} r_j M_{ji} h_i \,.
\end{align}

The next task is to determine the set of numerator polynomials $\{ {\cal N}_i\}_{i\in B}$ for the basis 
functions of all partial remainders from the random 
numerical evaluations of \cref{eqn:randomPS}.
Naively, determining ${\cal N}_i$ requires as many evaluations as there are 
free parameters in the polynomials ${\cal N}_i$. A large number of evaluations can be 
limiting due to the evaluation time of partial
remainders. Reducing the size of the required numerical sample is thus an important
goal.
Below we exploit two observations about the structure of the coefficient
functions $r_i$ to significantly reduce the size of the required numerical
samples: In section  \ref{sec:rescaling} we recycle the compact coefficients of
the two-loop five gluon amplitudes \cite{DeLaurentis:2023nss} and, in section
\ref{sec:funcSpaceConstr} we exploit linear basis changes to find new
coefficients in partial fractioned form. 

\subsection{Rescaled coefficient functions}
\label{sec:rescaling}

We construct a class of candidate coefficient functions from the known
functions for the five-gluon amplitudes \cite{DeLaurentis:2023nss}. To this end
we rescale the gluon coefficient functions by spinor brackets to match the
little-group weight of the quark amplitudes. This rescaling is inspired by analogous
factors in supersymmetry Ward identities which link tree amplitudes of gluon
and gluino states \cite{Grisaru:1977px,Grisaru:1976vm,Parke:1985pn} (see also
\cite{Elvang:2013cua}). Such a rescaling is expected from on-shell recursion
relations \cite{Britto:2005fq} for loop amplitudes
\cite{Bern:2005hh,Bern:2005hs,Bern:2005cq,Berger:2008sj} and collinear
factorization \cite{Bern:1994zx,Kosower:1999xi,Bern:1999ry} in general.  
Let us start from a gluon function from appendix C of
ref.~\cite{DeLaurentis:2023nss}, e.g.
\begin{equation}
\tilde{r}^{--}_{18}(1^{1},2^{1},3^{1},4^{-1},5^{-1}) = 
\frac{[12]\langle 24\rangle \langle 45\rangle }{\langle 12 \rangle \langle 23\rangle^2[25]} \, .
\end{equation}
First, in order to align the little-group weights with the two-quark-three gluon amplitudes,
we permute momentum labels
\begin{equation}
\tilde{r}^{--}_{18}(1^{1},4^{1},3^{1},5^{-1},2^{-1}) = 
\frac{[14] \langle 25\rangle \langle 45\rangle }{\langle 14\rangle [24]\langle 34\rangle^2}\,.
\end{equation}
Then, we can build functions for the process
$(u^{1/2},\bar{u}^{-1/2},g^1,g^1,g^{-1})$ 
by multiplying the gluon function by any function which raises and lowers
the little-group weights of legs 1 and 2 by one unit,
respectively. For instance, functions of the form
\begin{equation}\label{eqn:factor}
\frac{\langle 14\rangle }{\langle 24\rangle} \, ,
\end{equation}
correctly map the little-group weights.
The function that we obtain is
\begin{align}
\tilde{r}_{73}^-(1^{1/2},2^{-1/2},& 3^{1},4^{1},5^{-1}) 
= \frac{\langle 14 \rangle }{\langle 24\rangle }  
\frac{[14]\langle 25\rangle\langle 45\rangle}{\langle 14\rangle [24]\langle 34\rangle^2} \nonumber \\
& = \frac{[14] \langle 25\rangle\langle 45\rangle}{\langle 24\rangle [24]\langle 34\rangle^2} \, .
\end{align}
We then test numerically whether the resulting function belongs to the
vector space spanned by the coefficients of the $(u^{1/2},\bar{u}^{-1/2},g^1,g^1,g^{-1})$ 
MHV partial remainders,
\begin{align}
\vec {\tilde r}_i \in {\rm span} \{r_j \}_{j\in B}\,.
\end{align}
If it is in fact part of the span, we keep the function.
In this way we obtain a set of analytically known functions which allows to
parametrize part of the coefficient functions $r_i$.  We denote the set of
these functions by 
\begin{align}\label{eqn:trailBasis}
\{ \tilde r_i \}_{i\in \tilde B}
\end{align}
and index them by the label $i$ in the set $\tilde B$. To simplify the set, 
we remove linearly dependent functions $\tilde r_i$.

Let us note that we do not aim here to explore all possible rescaling factors.
For simplicity, we only consider factors such as that of \cref{eqn:factor} and
generalizations thereof with numerator and denominator mass dimensions not exceeding
two. 

After applying this rescaling procedure we obtain a significant number of coefficient
functions of the quark amplitudes from the gluon ones.
We obtain more than 50\% of the two-quark three-gluon
MHV functions by rescaling the five-gluon functions. 
Since the basis of two-quark three-gluon single minus functions can be reconstructed from a small number of sample evaluations,
we do not apply this strategy for them.
We also obtain the majority (more than 90\%) of the four-quark one-gluon
basis functions by rescaling two-quark three-gluon and the five-gluon
functions.

\subsection{Filling the space of coefficient functions}
\label{sec:funcSpaceConstr}

So far we have obtained a portion of the coefficient functions from
rescaling gluon coefficients, as discussed above in section \ref{sec:rescaling}. 
We will now reconstruct new rational functions which we add to the
set $\{ \tilde r_i \}_{i \in \tilde B}$ of \cref{eqn:trailBasis} until it spans the 
full function space $\{ r_i \}_{i \in B}$ of \cref{eqn:funcBasis}.
In order to simplify the discussion, we now assume that the sets $B$ and $\tilde B$ correspond
to a specific helicity class, i.e.~single minus or the MHV configuration.
By construction, the functions $\tilde r_i$ are in the linear span of the $r_i$. 
In terms of the vectors of function values, we have,
\begin{align}
   {\rm span} \{\vec{\tilde r}_i\}_{i\in \tilde B} \subseteq
   {\rm span} \{\vec r_i\}_{i\in B}\,. 
\end{align}
In the reconstruction of the missing functions, we now leverage the observation 
that partial fractioned coefficients 
take a simple form, with reduced mass dimension of numerator and denominator. 
We select the $r_i$ with lowest mass
dimension numerator, that does not lie in the span of $\{ \tilde r_i \}_{i\in
\tilde B}$. We then solve the partial fractioned ansatz \cite{Abreu:2023bdp}
\begin{equation}
r_i = \frac{\tilde{\mathcal{N}}_i}{\prod_j \mathcal{D}_j^{\tilde{q}_{ij}}} 
+ \sum_{k \in \tilde B} c_{ik} \tilde{r}_k \, . \label{eq:ansatz-removing-overlap}
\end{equation}
The parameters $\tilde{q}_{ij}$ in the ansatz are set empirically, i.e.~we
set a degree bound on the sum of powers $\sum_j\tilde{q}_{ij}\le q$. 
We then walk through all possible choices of
denominators $\prod_j \mathcal{D}_j^{\tilde{q}_{ij}}$ in \cref{eq:ansatz-removing-overlap} of the chosen
degree obtained as subsets of the original denominator
$\prod_j \mathcal{D}_j^{q_{ij}}$ in \cref{eqn:LCDimproved} and fit the numerator 
polynomial $\tilde {\cal N}_i$ and coefficients $c_{ik}$ using the
numerical sample evaluations from \cref{eqn:randomPS}. Given a successful fit, we then consider the function,
\begin{align}
\tilde r_i=\frac{\tilde{\mathcal{N}}_i}{\prod_j \mathcal{D}_j^{\tilde{q}_{ij}}} 
\end{align}
and all associated functions with permuted momentum labels 
and matching little-group weights. 
We add all such functions to the set of new, analytically known functions. 
We denote the updated set again with $\{ \tilde
r_j \}_{j \in \tilde B}$ with an adjusted index set $\tilde B$. 
The procedure is repeated until the linear span of the reconstructed functions 
covers the one of the coefficient functions,
\begin{align}\label{eq:span-supset}
   {\rm span} \{\vec{\tilde r}_i\}_{i\in \tilde B} \supseteq
   {\rm span} \{\vec r_i\}_{i\in B}\,. 
\end{align}
By construction, the only overshoot of the span in the left-hand side compared to
that in the right-hand side can be in the permutation closure of the generating
functions. No generating spinor-helicity function can be dropped while
keeping \cref{eq:span-supset} valid. Thus, the cost of this overshoot
is minimal, while potentially being helpful to further simplify the
basis.

We observe that this approach is very effective.
The degree bound of $q=12$ suffices for the analytic
reconstruction of all remaining coefficient functions from approximately $2k$
random numerical samples (\ref{eqn:randomPS}).

The method is efficient due to a number of implementational improvements: we
cache numerical values for both the right-hand side and the summation of
\cref{eq:ansatz-removing-overlap}. For each choice of $\tilde{q}_{ij}$, we only
need to re-generate an ansatz for $\mathcal{N}_i$, insert numerical values, and
perform a Gaussian elimination. For ans\"atze of this size, both construction
via \texttt{OR-tools} and row reduction via \texttt{linac} take
$\mathcal{O}(100\,\text{ms})$, meaning hundreds to thousands of guesses can be
checked in the time it takes to collect additional numerical samples for the
remainder functions.

\section{Results} 
\label{sec:results} 

We have obtained the analytic expression for the two-loop
two-quark three-gluon and four-quark one-gluon helicity finite remainders. The results are given in
ancillary files as explained in detail in section \ref{sec:ancillary}.

One of the central results of this work is the basis of rational coefficient functions $\tilde
r_i$, which we given in the appendix of this paper.  We label the four-quark one-gluon MHV functions
simply as $\tilde{r}$ and present them in \cref{sec:4q1g-mhv-basis}. The two-quark three gluon
functions are labelled based on the helicity of the last gluon, as
$\tilde{r}^{+}$ for the single minus configuration, and $\tilde{r}^{-}$ for the
MHV one.  We present them in presented in \cref{sec:2q3g-single-minus-basis}
and \cref{sec:2q3g-mhv-basis}, respectively. An account of the size of the
function bases is displayed in \cref{tab:VS-sizes}. 

\begin{table}[t]
\renewcommand{\arraystretch}{1.2}
\centering
\begin{tabular}{*{3}{C{15ex}}}
\toprule
Particle Helicities & Vector-space dimension & Generating set size \\
\colrule
$u^+\bar{u}^-g^+g^+\;\,g^+$ & 424 & 91 \\
$u^+\bar{u}^-g^+g^+\;\,g^-$ & 844 & 449 \\
\hline
$u^+\bar{u}^-d^+\bar{d}^-g^-$ & 435 & 124 \\
\botrule
\end{tabular}
\caption{
  \label{tab:VS-sizes} For each helicity configuration, this table
  shows the dimension of the vector space of rational functions, and
  the number of functions in the generating set that spans the space
  upon closure under the symmetries of the little-group weights. }
\end{table}

In order to facilitate future comparison with our results we collect reference
values for finite remainders in appendix \ref{sec:referenceEvaluations}.
Finally, in section \ref{sec:numEval} we present an efficient implementation of
our results and discuss its performance and stability.

\subsection{Validation}

In order to validate the amplitude computation we have performed a number of
checks. The evaluation of the amplitudes with the numerical unitarity method
was carried out with the well-tested program \textsc{Caravel} \cite{Abreu:2020xvt}.
At each phase-space point used in analytic reconstruction we check cancellation
of $\epsilon$ poles in finite remainders (\ref{eqn:finRemainder}).
To validate the analytic reconstruction, we check that the analytic results
match further evaluations in \textsc{Caravel} using finite-fields with a
different characteristic. 

Furthermore, we perform a number of checks on the hard function
(\ref{eq:hard-function}), which verifies the assembly of partial remainders based on
color and helicity sums.  We verified that all expected symmetries of the hard
functions (\ref{eq:hard-function}) under permutations of particles' momenta are
satisfied, as well the parity-conjugation symmetry.
We have compared our numerical evaluations of the one-loop hard functions
$\mathcal{H}^{(1)}$ (\ref{eq:hard-function-partial}) against \texttt{BlackHat}
\cite{Berger:2008sj} in all physical channels and found perfect agreement.
We have verified that upon taking the leading-color limit of
$\mathcal{H}^{(1)}$, $\mathcal{H}^{(2)}$,  we reproduce (after performing the
IR scheme change) the results resented in \cite{Abreu:2021oya}.
Finally, we have compared our results after performing the IR scheme change
(see \cref{sec:scheme-change}) with the numerical benchmarks presented in
\cite[Table II]{Agarwal:2023suw}. We find agreement with a revised
version of \cite{Agarwal:2023suw}.

\subsection{Numerical evaluation}
\label{sec:numEval}

We implement our analytic expressions for the generating set of helicity
partial remainders in \cref{eq:partial_2q,eq:partial_4q}, as well for the hard
functions in \cref{eq:hard-function,eq:hard-function-partial} for all physical
channels in \cref{eq:2q3g,eq:4q1g,eq:4q1g_identical} in a C++ library \cite{FivePointAmplitudes}.  Together with the
five-gluon channel from ref.~\cite{DeLaurentis:2023nss} this is the complete set
required for the calculation of NNLO QCD corrections for three jet production
at hadron colliders.

\begin{figure*}[ht]
  \begin{subfigure}{0.75\linewidth}
    \includegraphics[width=1\linewidth]{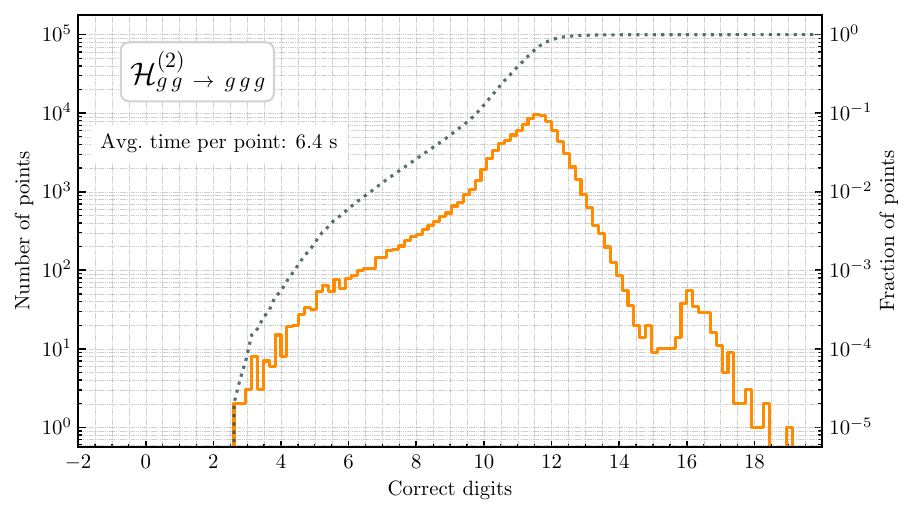}
  \end{subfigure}
  \begin{subfigure}{0.75\linewidth}
    \includegraphics[width=1\linewidth]{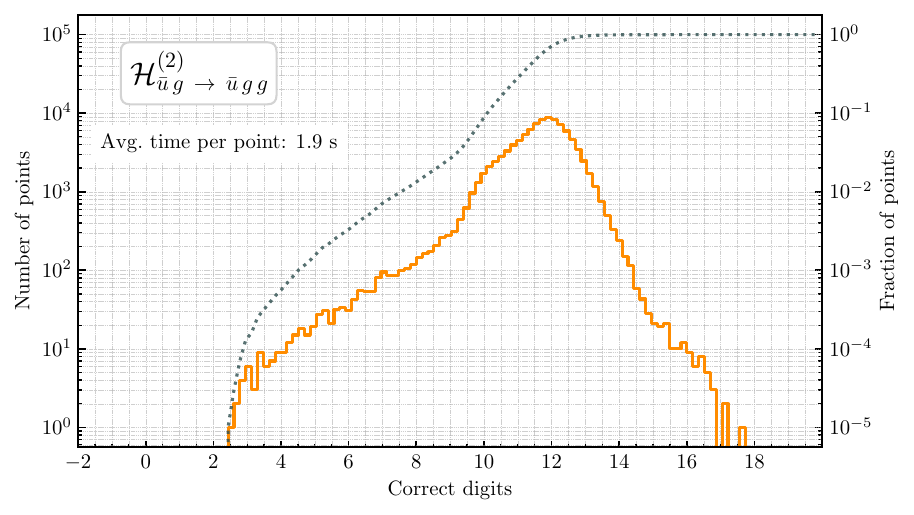}
  \end{subfigure}
  \begin{subfigure}{0.75\linewidth}
    \includegraphics[width=1\linewidth]{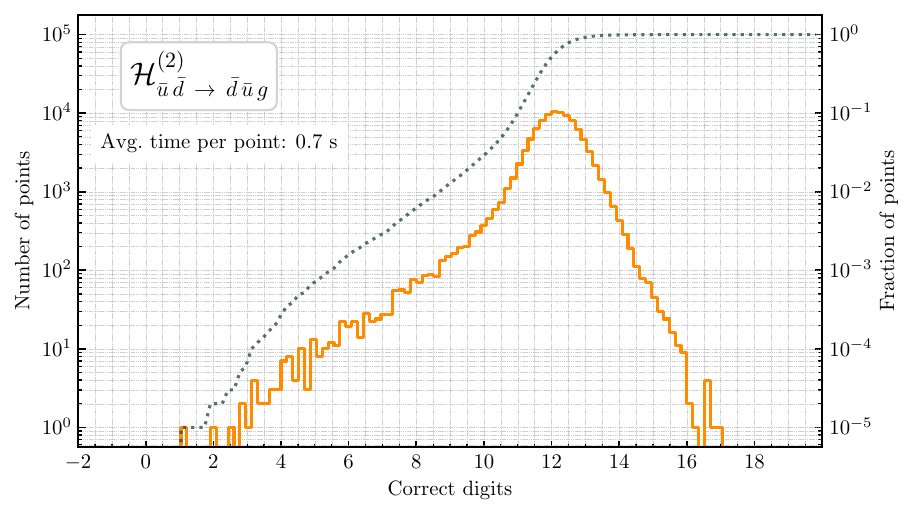}
  \end{subfigure}
  \caption{
    \label{fig:num_stab}
      Distributions of the base 10 logarithm of the relative error (correct
digits) of the NNLO hard functions defined in \cref{eq:hard-function} for
representative physical channels contributing to three-jet production at hadron
colliders.  The dashed curves represent respective cumulative distributions.
Here $N_f$ is set to $5$, and the renormalization scale is set dynamically to
(half of) the sum of the transverse momentum of the final-state partons.  
The phase space definition is taken from ref.~\cite{ATLAS:2014qmg}.
  }
\end{figure*}

To demonstrate the numerical performance of our implementation, we sample three
representative channels over 100k points from the phase space of 
ref.~\cite{ATLAS:2014qmg} (see also \cite{Abreu:2021oya}).  For completeness, we
also study the five-gluon channel calculated in \cite{DeLaurentis:2023nss}.
The evaluations are compared to target values computed in quadruple precision, and the
distribution of the base 10 logarithm of the relative error (correct digits) is
shown in \cref{fig:num_stab} (c.f.~ref.~\cite{Abreu:2021oya}).
We observe that despite the markedly increased
complexity of subleading color amplitudes, the numerical performance of our
results is excellent and comparable to the leading-color results reported in
\cite{Abreu:2021oya}.  The rescue system developed in \cite{Abreu:2021oya} is
effective in capturing unstable points. 
This is especially relevant for the five-gluon channel where we observe the second small peak in the distribution at about 16 digits
formed by the phase-space points rescued through a quadruple precision evaluation.

We observe (in the panels of \cref{fig:num_stab}) average evaluation times per phase-space point of few seconds in all
production channels, already taking into account the time required to detect
and rescue unstable points in quadruple precision. 
In contrast to the leading-color approximation, where most of the evaluation
time is spent on pentagon functions, the evaluation in full color is dominated
by the contraction of indices in \cref{eq:coeffBasis}.  This hints that further
improvements are achievable if a more tailored basis of transcendental
functions is used.  
The observed
stability and evaluation times will enable seamless lifting of the
leading-color approximation which has been employed in cross-section
computations so far \cite{Czakon:2021mjy,Chen:2022ktf}.

\vspace{-1mm}
\subsection{Ancillary files}
\label{sec:ancillary}
We provide ancillary files for all independent partial finite
remainders, including all crossing \cite{de_laurentis_2023_10227002}. We organize the ancillay files in
the same manner as ref.~\cite{DeLaurentis:2023nss}. Overall, for the
complete five-parton computation, the folder structure is as follows:
\begin{enumerate}
\item[] \texttt{ggggg/}
\item[] $\qquad$\texttt{all\_plus/}
\item[] $\qquad$\texttt{single\_minus/}
\item[] $\qquad$\texttt{mhv/}
\item[] \texttt{uubggg/}
\item[] $\qquad$\texttt{single\_minus/}
\item[] $\qquad$\texttt{mhv/}
\item[] \texttt{uubddbg/}
\item[] $\qquad$\texttt{mhv/}
\end{enumerate}
For each of these folders, representing external states and the
associated helicity configuration, we provide the bases $h_j$ and
$\tilde{r}_i$ (\ref{eq:coeffBasis}), respectively in the files
\begin{enumerate}
\item \texttt{basis\_transcendental}$\,,$
\item \texttt{basis\_rational}$\,.$
\end{enumerate}
Further subfolders contain the matrices $M_{ij}$ of rational
numbers (\ref{eq:coeffBasis}). The folders are labelled as
\begin{enumerate}
\item[] \texttt{\{}$\vec h$\texttt{\}\_\{}$L$\texttt{\}L\_Nc\{}$n_c$\texttt{\}\_Nf\{}$n_f$\texttt{\}/}$\,,$
\end{enumerate}
where $\vec h$, $L$, $n_c$ and $n_f$ refer to helicities, number of
loops, number of $N_c$ powers and number of $N_f$ powers. Since $L$,
$n_c$ and $n_f$ do not always identify a unique partial, for $A_2$ and
$A_3$ we extend this notation to
\begin{enumerate}
\item[] \texttt{\{}$\vec h$\texttt{\}\_\{}$L$\texttt{\}L\_Nc\{}$n_c$\texttt{\}\_Nf\{}$n_f$\texttt{\}\_\{}integers\texttt{\}/}$\,,$
\end{enumerate}
where the extra “integers” represent the split in the fundamental
generators, i.e.~\texttt{\_2\_1} for $A_2$ and \texttt{\_3\_0} for
$A_3$. The rational matrices are named as
\begin{enumerate}
\item[] \texttt{rational\_matrix\_\{permutation\}}$\,,$
\end{enumerate}
for each permutation of the external legs, which may involve
crossings.

We further provide assembly scripts within the ancillary files.

\FloatBarrier

\section{Conclusions}

We have presented the computation of the two-quark
three-gluon and four-quark one-gluon amplitudes at two loops in
QCD. We derive compact analytic expressions in the spinor-helicity
formalism for the finite remainders, including all contributions
beyond the leading-color approximation and all crossings. 

We systematically investigate linear relations among partial remainders in the trace
basis of the color generators, relying on numerical amplitude evaluation and
linear algebra. We do not find any non-trivial
identities among the two-loop two-quark three-gluon partial remainders, while we
obtain six identities among two-loop four-quark one-gluon partial remainders.

With regards to the analytic reconstruction, we explore a new method to obtain
quark amplitudes from gluon ones, inspired by supersymmetry Ward identities.
This entails rescaling the gluon spinor-helicity basis functions
presented in \cite{DeLaurentis:2023nss} by simple factors carrying the
appropriate little-group weights, such as ratios of the respective tree
amplitudes.  

Finally, we provide the efficient C++ code \cite{FivePointAmplitudes} for the
computation of color- and helicity-summed squared matrix elements,
suitable for immediate phenomenological applications.

Together with our earlier results of the compact gluon amplitudes
\cite{DeLaurentis:2023nss} this completes our computation of the two-loop
five-parton amplitudes in full color. 
We envisage their application to both new
phenomenological studies, as well as novel theoretical investigations into the
perturbative structure of QCD.

\begin{acknowledgements}

We gratefully acknowledge discussions with Samuel Abreu and Ben Page.
We thank Maximillian Klinkert for collaboration during the initial stages of this work.
We thank Bakul Agarwal, Federico Buccioni, Federica Devoto,
Giulio Gambuti, Andreas von Manteuffel, Lorenzo Tancredi for the
numerical comparison of reference values for the two-loop five-parton 
QCD amplitudes.
We thank Matteo Marcoli for discussions about the tripole correlation terms in Catani's scheme.
We gratefully acknowledge the computing resources provided by the Paul
Scherrer Insitut (PSI) and the University of Zurich (UZH).
V.S.\ has received funding from the European Research Council (ERC)
under the European Union's Horizon 2020 research and innovation
programme grant agreement 101019620 (ERC Advanced Grant TOPUP).
G.D.L.'s work is supported in part by the U.K.\ Royal Society through
Grant URF\textbackslash R1\textbackslash 20109.
\end{acknowledgements}

\appendix

\onecolumngrid
\newgeometry{left=1in, right=1in, top=1in, bottom=1in}  

\section{Reference evaluations}
\label{sec:referenceEvaluations}

We present in \cref{tab:target} the reference evaluations of the hard functions
defined in \cref{eq:hard-function,eq:hard-function-partial} at the point
\begin{equation} \label{eq:referencePointBis}
  \begin{aligned}
    p_1 &= \{-3.6033749869055013, 3.5549594933215615, 0.033937560795432568, 0.58772658529828721\},\\
    p_2 &= \{-3.5779991067160259, 3.5333697062718894, 0.033731453043168395, -0.56235070510881185\},\\
    p_3 &= \{1.7967619455543639, -1.7454551820128482, 0.10779867188908804, -0.41245501926361226\},\\
    p_4 &= \{0.41554983516150722, -0.38259279840807596, -0.13137170294301400, 0.095109893149375643\},\\
    p_5 &= \{4.9690623129056561, -4.9602812191725268, -0.044095982784675001, 0.29196924592476125\},
  \end{aligned}
\end{equation}
with the renormalization scale set to $\mu=1$.  For comparison we also show in
\cref{tab:target-lc} the evaluations of the same hard functions in the leading
color approximation.  The evaluations are produced using the numerical code
\cite{FivePointAmplitudes} which we make available with this work.  We remind
the reader that whenever we are using the $\to$ notation, the first two
particles are to be understood as crossed to be incoming.

\begin{table}[!h]
  \renewcommand{\arraystretch}{1.4}
  \centering
\begin{tabular}{c*{5}{C{20ex}}}
    \toprule
    Channel&$H^{(1)[0]}$&$H^{(1)[1]}$&$H^{(2)[0]}$&$H^{(2)[1]}$&$H^{(2)[2]}$ \\
	\colrule
${\bar u}u\rightarrow ggg$ & $-0.006212852706$ & $-0.0454101007307$ & $318.955641322$ & $-14.6420604152$ & $0.0733824569231$ \\
${\bar u}g\rightarrow {\bar u}gg$ & $39.0994013942$ & $-0.492481498134$ & $2066.36392385$ & $-114.543895236$ & $0.861580772061$ \\
$gg\rightarrow u{\bar u}g$ & $4.07662153685$ & $0.0130905622339$ & $463.236973933$ & $-23.8189711025$ & $-0.430932958459$ \\
	\colrule
${\bar u}u\rightarrow d{\bar d}g$ & $17.8776395916$ & $-1.83734127116$ & $552.343071009$ & $-60.0556784182$ & $1.52922442321$ \\
${\bar u}d\rightarrow d{\bar u}g$ & $45.4231876826$ & $-1.80612761041$ & $1876.8079204$ & $-158.012839383$ & $2.48406129185$ \\
${\bar u}{\bar d}\rightarrow {\bar d}{\bar u}g$ & $36.5463358316$ & $-2.14821130255$ & $1545.41004188$ & $-158.008191002$ & $3.32697171361$ \\
${\bar u}g\rightarrow {\bar u}d{\bar d}$ & $56.1144160759$ & $-2.88325090618$ & $2879.870086$ & $-278.230078768$ & $6.18882069862$ \\
	\colrule
${\bar u}u\rightarrow u{\bar u}g$ & $35.1251353834$ & $-1.81972845991$ & $1384.61457589$ & $-121.556173952$ & $2.13373315494$ \\
${\bar u}{\bar u}\rightarrow {\bar u}{\bar u}g$ & $71.4608675873$ & $-3.67999937996$ & $4392.67315319$ & $-441.108685487$ & $9.7657279991$ \\
${\bar u}g\rightarrow {\bar u}u{\bar u}$ & $54.3760288328$ & $-2.83165571182$ & $2801.23577819$ & $-269.746150468$ & $5.98133163379$ \\
    \botrule
  \end{tabular}
  \caption{
   \label{tab:target}
   Reference values for hard functions of each channel considered in this work,
as defined in \cref{eq:hard-function-partial}.  Here we set $N_c = 3$.
  }
\end{table}

\begin{table}[!h]
  \renewcommand{\arraystretch}{1.4}
  \centering
  \begin{tabular}{c*{5}{C{20ex}}}
    \toprule
    Channel&$H^{(1)[0]}$&$H^{(1)[1]}$&$H^{(2)[0]}$&$H^{(2)[1]}$&$H^{(2)[2]}$ \\
	\colrule
${\bar u}u\rightarrow ggg$ & $-0.904169945632$ & $-0.0290745413125$ & $286.504396459$ & $-16.4877211236$ & $0.0591610370825$ \\
${\bar u}g\rightarrow {\bar u}gg$ & $35.3138994216$ & $-0.491121881165$ & $1709.35891402$ & $-105.715888821$ & $0.872329205733$ \\
$gg\rightarrow u{\bar u}g$ & $3.62839265602$ & $0.0106464749312$ & $398.255787625$ & $-23.2749530656$ & $-0.380163822808$ \\
	\colrule
${\bar u}u\rightarrow d{\bar d}g$ & $19.4220143259$ & $-1.78898228361$ & $568.84899353$ & $-62.8383318838$ & $1.40770779745$ \\
${\bar u}d\rightarrow d{\bar u}g$ & $45.7223617721$ & $-1.73638756796$ & $1852.5688152$ & $-151.61506626$ & $2.31111445374$ \\
${\bar u}{\bar d}\rightarrow {\bar d}{\bar u}g$ & $26.2821708156$ & $-2.3018897729$ & $1159.19990841$ & $-139.532378448$ & $3.70244721968$ \\
${\bar u}g\rightarrow {\bar u}d{\bar d}$ & $50.9489640028$ & $-2.91394709477$ & $2497.22216028$ & $-263.212680843$ & $6.36125177509$ \\
	\colrule
${\bar u}u\rightarrow u{\bar u}g$ & $35.8324660681$ & $-1.75616511096$ & $1369.84305747$ & $-118.231756564$ & $1.9714004014$ \\
${\bar u}{\bar u}\rightarrow {\bar u}{\bar u}g$ & $66.9539031696$ & $-3.68077608088$ & $3941.34122482$ & $-416.819803981$ & $9.78316003718$ \\
${\bar u}g\rightarrow {\bar u}u{\bar u}$ & $48.3172172347$ & $-2.84429678497$ & $2375.45809856$ & $-250.314601734$ & $6.06149093974$ \\
    \botrule
  \end{tabular}
  \caption{
   \label{tab:target-lc}
   Reference values for hard functions in the leading color approximation of
each channel considered in this work, as defined in
\cref{eq:hard-function-partial}.  Here we set $N_c = 3$.
  }
\end{table}

\newgeometry{left=0.6in, right=0.8in, top=0.8in, bottom=0.8in}

\section{Four-quark one-gluon MHV functions}
\label{sec:4q1g-mhv-basis}

\vspace{-6mm}
\hskip-0.4cm
\begin{tabular}{p{4.3cm}@{}p{4.5cm}@{}p{4.7cm}@{}p{4.7cm}}
\begin{equation}\begin{gathered}
\small \tilde{r}_{1} = \frac{⟨24⟩}{⟨15⟩⟨35⟩}
\end{gathered}\hspace{-2mm} \nonumber 
\end{equation}
&
\begin{equation}\begin{gathered}
\small \tilde{r}_{20} = \frac{[35][25]}{⟨13⟩[23][24]}
\end{gathered}\hspace{-2mm} \nonumber 
\end{equation}
&
\begin{equation}\begin{gathered}
\small \tilde{r}_{39} = \frac{⟨23⟩^2[23]}{⟨12⟩[24]⟨35⟩^2}
\end{gathered}\hspace{-2mm} \nonumber 
\end{equation}
&
\begin{equation}\begin{gathered}
\small \tilde{r}_{58} = \frac{[13]⟨24⟩⟨14⟩}{⟨13⟩[34]⟨45⟩^2}
\end{gathered}\hspace{-2mm} \nonumber 
\end{equation}
\\[-8.8mm]
\begin{equation}\begin{gathered}
\small \tilde{r}_{2} = \frac{[15]}{[24]⟨35⟩}
\end{gathered}\hspace{-2mm} \nonumber 
\end{equation}
&
\begin{equation}\begin{gathered}
\small \tilde{r}_{21} = \frac{[13][25]}{[23][24]⟨35⟩}
\end{gathered}\hspace{-2mm} \nonumber 
\end{equation}
&
\begin{equation}\begin{gathered}
\small \tilde{r}_{40} = \frac{⟨45⟩[23][15]}{[12]^2⟨15⟩^2}
\end{gathered}\hspace{-2mm} \nonumber 
\end{equation}
&
\begin{equation}\begin{gathered}
\small \tilde{r}_{59} = \frac{[35][13]⟨23⟩}{⟨34⟩[34]^2⟨35⟩}
\end{gathered}\hspace{-2mm} \nonumber 
\end{equation}
\\[-8.8mm]
\begin{equation}\begin{gathered}
\small \tilde{r}_{3} = \frac{⟨23⟩⟨24⟩}{⟨12⟩⟨35⟩^2}
\end{gathered}\hspace{-2mm} \nonumber 
\end{equation}
&
\begin{equation}\begin{gathered}
\small \tilde{r}_{22} = \frac{[35]⟨23⟩}{⟨15⟩⟨35⟩[45]}
\end{gathered}\hspace{-2mm} \nonumber 
\end{equation}
&
\begin{equation}\begin{gathered}
\small \tilde{r}_{41} = \frac{[35]^2⟨45⟩}{⟨12⟩[23]^2⟨35⟩}
\end{gathered}\hspace{-2mm} \nonumber 
\end{equation}
&
\begin{equation}\begin{gathered}
\small \tilde{r}_{60} = \frac{[23]⟨24⟩⟨12⟩}{⟨15⟩^2⟨25⟩[25]}
\end{gathered}\hspace{-2mm} \nonumber 
\end{equation}
\\
\end{tabular}

\pagebreak

\hskip-0.3cm
\begin{tabular}{p{5.6cm}@{}p{5.7cm}@{}p{6.3cm}}
\begin{equation}\begin{gathered}
\small \tilde{r}_{77} = \frac{⟨25⟩^2[15]^2}{⟨12⟩[14]^2⟨35⟩⟨45⟩}
\end{gathered}\hspace{-2mm} \nonumber 
\end{equation}
&
\begin{equation}\begin{gathered}
\small \tilde{r}_{96} = \frac{[35]^2[12]⟨12⟩}{⟨15⟩⟨25⟩[25]^2[34]}
\end{gathered}\hspace{-2mm} \nonumber 
\end{equation}
&
\begin{equation}\begin{gathered}
\small \tilde{r}_{115} = \frac{[34]⟨12⟩[23]⟨34⟩⟨45⟩}{[12]⟨15⟩^3⟨25⟩[25]}
\end{gathered}\hspace{-2mm} \nonumber 
\end{equation}
\\[-8.5mm]
\begin{equation}\begin{gathered}
\small \tilde{r}_{78} = \frac{[13]^2⟨23⟩^2}{⟨12⟩[14]^2⟨35⟩⟨45⟩}
\end{gathered}\hspace{-2mm} \nonumber 
\end{equation}
&
\begin{equation}\begin{gathered}
\small \tilde{r}_{97} = \frac{[35]⟨13⟩⟨24⟩^2}{⟨12⟩^2[25]⟨35⟩⟨45⟩}
\end{gathered}\hspace{-2mm} \nonumber 
\end{equation}
&
\begin{equation}\begin{gathered}
\small \tilde{r}_{116} = \frac{⟨25⟩[24][35]⟨45⟩⟨23⟩}{⟨15⟩[34]⟨35⟩^2⟨5|1\tiny{+}2|5]}
\end{gathered}\hspace{-2mm} \nonumber 
\end{equation}
\\[-8.5mm]
\begin{equation}\begin{gathered}
\small \tilde{r}_{79} = \frac{[25]^2⟨25⟩^2}{⟨12⟩[24]^2⟨35⟩⟨45⟩}
\end{gathered}\hspace{-2mm} \nonumber 
\end{equation}
&
\begin{equation}\begin{gathered}
\small \tilde{r}_{98} = \frac{[25]^2⟨23⟩⟨12⟩}{⟨13⟩^2[24]⟨2|1\tiny{+}3|2]}
\end{gathered}\hspace{-2mm} \nonumber 
\end{equation}
&
\begin{equation}\hspace{-5mm}\begin{gathered}
\small \tilde{r}_{117} = \frac{[35]^3⟨45⟩⟨25⟩⟨13⟩}{⟨12⟩⟨14⟩[23]⟨34⟩[34]^2⟨35⟩}
\end{gathered}\hspace{-5mm} \nonumber 
\end{equation}
\\
\end{tabular}

\newgeometry{left=0.6in, right=0.8in, top=0.8in, bottom=0.8in}

\section{Two-quark three-gluon single-minus functions}
\label{sec:2q3g-single-minus-basis}

\vspace{-6mm}
\hskip-0.2cm
\begin{tabular}{p{5.1cm}@{}p{5.8cm}@{}p{6.4cm}}
\begin{equation}\begin{gathered}
\small \tilde{r}^{+}_{1} = \frac{[15]^2}{[12]⟨34⟩^2}
\end{gathered}\hspace{-2mm} \nonumber 
\end{equation}
&
\begin{equation}\begin{gathered}
\small \tilde{r}^{+}_{20} = \frac{[34][15]⟨25⟩}{[14]⟨15⟩⟨34⟩⟨45⟩}
\end{gathered}\hspace{-2mm} \nonumber 
\end{equation}
&
\begin{equation}\begin{gathered}
\small \tilde{r}^{+}_{39} = \frac{⟨25⟩[12]⟨23⟩[35]}{[23]⟨24⟩⟨34⟩⟨35⟩^2}
\end{gathered}\hspace{-2mm} \nonumber 
\end{equation}
\\[-9mm]
\begin{equation}\begin{gathered}
\small \tilde{r}^{+}_{2} = \frac{⟨12⟩[15]}{⟨15⟩⟨34⟩^2}
\end{gathered}\hspace{-2mm} \nonumber 
\end{equation}
&
\begin{equation}\begin{gathered}
\small \tilde{r}^{+}_{21} = \frac{[45][34]⟨24⟩}{⟨15⟩⟨34⟩⟨35⟩[35]}
\end{gathered}\hspace{-2mm} \nonumber 
\end{equation}
&
\begin{equation}\begin{gathered}
\small \tilde{r}^{+}_{40} = \frac{[45]⟨24⟩⟨12⟩[15]}{⟨14⟩^2[14]⟨23⟩⟨34⟩}
\end{gathered}\hspace{-2mm} \nonumber 
\end{equation}
\\[-9mm]
\begin{equation}\begin{gathered}
\small \tilde{r}^{+}_{3} = \frac{⟨23⟩[14]}{⟨34⟩⟨35⟩^2}
\end{gathered}\hspace{-2mm} \nonumber 
\end{equation}
&
\begin{equation}\begin{gathered}
\small \tilde{r}^{+}_{22} = \frac{[34][23]⟨12⟩}{⟨14⟩⟨15⟩[24]⟨45⟩}
\end{gathered}\hspace{-2mm} \nonumber 
\end{equation}
&
\begin{equation}\begin{gathered}
\small \tilde{r}^{+}_{41} = \frac{[13][34]⟨14⟩⟨23⟩}{⟨13⟩⟨45⟩^2⟨4|1\tiny{+}5|4]}
\end{gathered}\hspace{-2mm} \nonumber 
\end{equation}
\\
\end{tabular}

\pagebreak

\hskip-0.2cm
\begin{tabular}{p{5.9cm}@{}p{6.1cm}@{}p{5.9cm}}
\begin{equation}\begin{gathered}
\small \tilde{r}^{+}_{58} = \frac{⟨3|1\tiny{+}5|3]⟨24⟩^2⟨23⟩[25]}{⟨15⟩⟨25⟩⟨34⟩^4[35]}
\end{gathered}\hspace{-2mm} \nonumber 
\end{equation}
&
\begin{equation}\begin{gathered}
\small \tilde{r}^{+}_{74} = \frac{[25]⟨25⟩^2[12]⟨15⟩⟨24⟩^2}{⟨12⟩⟨13⟩[15]⟨35⟩⟨45⟩^4}
\end{gathered}\hspace{-2mm} \nonumber 
\end{equation}
&
\begin{equation}\begin{gathered}
\small \tilde{r}^{+}_{86} = \frac{-[45]⟨14⟩[13]⟨12⟩[14]}{⟨13⟩⟨45⟩⟨1|2\tiny{+}3|1]^2}+\\
\small\frac{-2[45]⟨12⟩⟨15⟩[15]}{⟨13⟩^2⟨45⟩⟨1|2\tiny{+}3|1]}\phantom{+}
\end{gathered}\hspace{-2mm} \nonumber 
\end{equation}
\\[-19mm]
\begin{equation}\begin{gathered}
\small \tilde{r}^{+}_{59} = \frac{⟨3|1\tiny{+}2|3][12]⟨12⟩^2}{⟨13⟩^2[13]⟨15⟩⟨34⟩⟨45⟩}
\end{gathered}\hspace{-2mm} \nonumber 
\end{equation}
&
\begin{equation}\hspace{-2mm}\begin{gathered}
\small \tilde{r}^{+}_{75} = \frac{\text{tr}_5^2}{⟨14⟩⟨15⟩⟨34⟩⟨35⟩[12]s_{13}s_{23}}
\end{gathered}\hspace{-3mm} \nonumber 
\end{equation}
&
\vspace{6mm}
\begin{equation}\begin{gathered}
\small \tilde{r}^{+}_{87} = \frac{[13][35][14]}{[12]⟨45⟩⟨3|1\tiny{+}2|3]}+\\
\small\frac{2[13]^2[45][34]⟨34⟩}{[12]⟨45⟩⟨3|1\tiny{+}2|3]^2}+\\
\small(12345\; \rightarrow \; 12543)\phantom{+}
\end{gathered}\hspace{-2mm} \nonumber 
\end{equation}
\\
\end{tabular}

\newgeometry{left=0.6in, right=0.8in, top=0.8in, bottom=0.8in}

\section{Two-quark three-gluon MHV functions}
\label{sec:2q3g-mhv-basis}

\vspace{-8mm}
\hskip-0.4cm
\begin{tabular}{p{4.5cm}@{}p{4.5cm}@{}p{4.5cm}@{}p{4.5cm}}
\begin{equation}\begin{gathered}
\small \tilde{r}^{-}_{1} = \frac{⟨25⟩^2}{⟨12⟩⟨34⟩^2}
\end{gathered}\hspace{-2mm}\nonumber
\end{equation}
&
\begin{equation}\begin{gathered}
\small \tilde{r}^{-}_{21} = \frac{⟨25⟩[34]}{⟨14⟩⟨34⟩[45]}
\end{gathered}\hspace{-2mm}\nonumber
\end{equation}
&
\begin{equation}\begin{gathered}
\small \tilde{r}^{-}_{41} = \frac{[14]⟨45⟩^2}{[12]⟨14⟩⟨34⟩^2}
\end{gathered}\hspace{-2mm}\nonumber
\end{equation}
&
\begin{equation}\begin{gathered}
\small \tilde{r}^{-}_{61} = \frac{[14]^2⟨25⟩^2}{⟨12⟩[15]^2⟨35⟩^2}
\end{gathered}\hspace{-2mm}\nonumber
\end{equation}
\\[-9.5mm]
\begin{equation}\begin{gathered}
\small \tilde{r}^{-}_{2} = \frac{⟨25⟩[12]}{[25]⟨34⟩^2}
\end{gathered}\hspace{-2mm}\nonumber
\end{equation}
&
\begin{equation}\begin{gathered}
\small \tilde{r}^{-}_{22} = \frac{⟨25⟩[34]}{⟨14⟩⟨23⟩[25]}
\end{gathered}\hspace{-2mm}\nonumber
\end{equation}
&
\begin{equation}\begin{gathered}
\small \tilde{r}^{-}_{42} = \frac{⟨25⟩[34]^2}{⟨13⟩⟨35⟩[35]^2}
\end{gathered}\hspace{-2mm}\nonumber
\end{equation}
&
\begin{equation}\begin{gathered}
\small \tilde{r}^{-}_{62} = \frac{[13]^2[14]^2}{⟨12⟩[12]^2[15]^2}
\end{gathered}\hspace{-2mm}\nonumber
\end{equation}
\\[-9.5mm]
\begin{equation}\begin{gathered}
\small \tilde{r}^{-}_{3} = \frac{⟨25⟩^2}{⟨13⟩⟨24⟩⟨34⟩}
\end{gathered}\hspace{-2mm}\nonumber
\end{equation}
&
\begin{equation}\begin{gathered}
\small \tilde{r}^{-}_{23} = \frac{[34][13]}{[25]⟨34⟩[35]}
\end{gathered}\hspace{-2mm}\nonumber
\end{equation}
&
\begin{equation}\begin{gathered}
\small \tilde{r}^{-}_{43} = \frac{⟨25⟩[14]^2}{⟨13⟩[15]^2⟨35⟩}
\end{gathered}\hspace{-2mm}\nonumber
\end{equation}
&
\begin{equation}\begin{gathered}
\small \tilde{r}^{-}_{63} = \frac{⟨12⟩^3[13]^2}{⟨14⟩^2⟨23⟩^2[35]^2}
\end{gathered}\hspace{-2mm}\nonumber
\end{equation}
\\
\end{tabular}

\pagebreak

\hskip-0.4cm
\begin{tabular}{p{4.5cm}@{}p{4.5cm}@{}p{4.5cm}@{}p{4.5cm}}
\begin{equation}\begin{gathered}
\small \tilde{r}^{-}_{81} = \frac{⟨25⟩⟨45⟩[13]}{[23]⟨24⟩⟨34⟩^2}
\end{gathered}\hspace{-2mm}\nonumber
\end{equation}
&
\begin{equation}\begin{gathered}
\small \tilde{r}^{-}_{101} = \frac{[23]⟨25⟩^2}{⟨14⟩⟨24⟩[25]⟨35⟩}
\end{gathered}\hspace{-2mm}\nonumber
\end{equation}
&
\begin{equation}\begin{gathered}
\small \tilde{r}^{-}_{121} = \frac{[14]⟨23⟩⟨15⟩^2}{⟨13⟩^3[15]⟨45⟩}
\end{gathered}\hspace{-2mm}\nonumber
\end{equation}
&
\begin{equation}\begin{gathered}
\small \tilde{r}^{-}_{141} = \frac{⟨45⟩^2[14]^2}{[12]^2⟨13⟩⟨24⟩⟨34⟩}
\end{gathered}\hspace{-2mm}\nonumber
\end{equation}
\\[-9.5mm]
\begin{equation}\begin{gathered}
\small \tilde{r}^{-}_{82} = \frac{⟨25⟩[34]⟨15⟩}{⟨14⟩^2⟨23⟩[24]}
\end{gathered}\hspace{-2mm}\nonumber
\end{equation}
&
\begin{equation}\begin{gathered}
\small \tilde{r}^{-}_{102} = \frac{[14]^2⟨15⟩}{[25]⟨34⟩⟨35⟩[45]}
\end{gathered}\hspace{-2mm}\nonumber
\end{equation}
&
\begin{equation}\begin{gathered}
\small \tilde{r}^{-}_{122} = \frac{[14]⟨25⟩^3}{⟨23⟩⟨24⟩^2[24]⟨35⟩}
\end{gathered}\hspace{-2mm}\nonumber
\end{equation}
&
\begin{equation}\begin{gathered}
\small \tilde{r}^{-}_{142} = \frac{[34]^2⟨35⟩⟨15⟩}{⟨13⟩^2⟨23⟩[23]^2}
\end{gathered}\hspace{-2mm}\nonumber
\end{equation}
\\[-9.5mm]
\begin{equation}\begin{gathered}
\small \tilde{r}^{-}_{83} = \frac{[14]⟨35⟩⟨24⟩}{⟨23⟩[25]⟨34⟩^2}
\end{gathered}\hspace{-2mm}\nonumber
\end{equation}
&
\begin{equation}\begin{gathered}
\small \tilde{r}^{-}_{103} = \frac{⟨45⟩[14]^2}{[15][25]⟨34⟩⟨35⟩}
\end{gathered}\hspace{-2mm}\nonumber
\end{equation}
&
\begin{equation}\begin{gathered}
\small \tilde{r}^{-}_{123} = \frac{⟨35⟩^2⟨12⟩[45]}{⟨13⟩^3[15]⟨34⟩}
\end{gathered}\hspace{-2mm}\nonumber
\end{equation}
&
\begin{equation}\begin{gathered}
\small \tilde{r}^{-}_{143} = \frac{[34]^2⟨25⟩^2}{⟨15⟩⟨23⟩⟨35⟩[35]^2}
\end{gathered}\hspace{-2mm}\nonumber
\end{equation}
\\
\end{tabular}

\pagebreak

\hskip-0.4cm
\begin{tabular}{p{4.5cm}@{}p{4.5cm}@{}p{4.5cm}@{}p{4.5cm}}
\begin{equation}\begin{gathered}
\small \tilde{r}^{-}_{161} = \frac{⟨45⟩^2⟨13⟩[14]^2}{⟨24⟩[24]^2⟨34⟩^3}
\end{gathered}\hspace{-2mm}\nonumber
\end{equation}
&
\begin{equation}\begin{gathered}
\small \tilde{r}^{-}_{181} = \frac{[13]^2⟨35⟩⟨45⟩}{[12]^2⟨14⟩⟨24⟩⟨34⟩}
\end{gathered}\hspace{-2mm}\nonumber
\end{equation}
&
\begin{equation}\begin{gathered}
\small \tilde{r}^{-}_{201} = \frac{[34]^2⟨45⟩⟨35⟩}{⟨13⟩⟨23⟩[23]^2⟨34⟩}
\end{gathered}\hspace{-2mm}\nonumber
\end{equation}
&
\begin{equation}\begin{gathered}
\small \tilde{r}^{-}_{221} = \frac{⟨34⟩⟨25⟩^3⟨15⟩}{⟨14⟩^2⟨23⟩⟨24⟩⟨35⟩^2}
\end{gathered}\hspace{-2mm}\nonumber
\end{equation}
\\[-9.5mm]
\begin{equation}\begin{gathered}
\small \tilde{r}^{-}_{162} = \frac{⟨25⟩⟨35⟩[14]}{[12]⟨13⟩⟨23⟩⟨34⟩}
\end{gathered}\hspace{-2mm}\nonumber
\end{equation}
&
\begin{equation}\begin{gathered}
\small \tilde{r}^{-}_{182} = \frac{⟨24⟩^2[34]⟨15⟩}{⟨14⟩^2⟨23⟩⟨34⟩[35]}
\end{gathered}\hspace{-2mm}\nonumber
\end{equation}
&
\begin{equation}\begin{gathered}
\small \tilde{r}^{-}_{202} = \frac{[34]^2⟨45⟩⟨25⟩}{⟨12⟩⟨23⟩[23]^2⟨34⟩}
\end{gathered}\hspace{-2mm}\nonumber
\end{equation}
&
\begin{equation}\begin{gathered}
\small \tilde{r}^{-}_{222} = \frac{[15]^2⟨15⟩^3}{⟨12⟩⟨14⟩[25]^2⟨34⟩⟨35⟩}
\end{gathered}\hspace{-2mm}\nonumber
\end{equation}
\\[-9.5mm]
\begin{equation}\begin{gathered}
\small \tilde{r}^{-}_{163} = \frac{⟨25⟩⟨12⟩[34]}{⟨13⟩⟨14⟩⟨23⟩[35]}
\end{gathered}\hspace{-2mm}\nonumber
\end{equation}
&
\begin{equation}\begin{gathered}
\small \tilde{r}^{-}_{183} = \frac{[13]⟨12⟩⟨15⟩^2}{⟨13⟩⟨14⟩^2⟨23⟩[23]}
\end{gathered}\hspace{-2mm}\nonumber
\end{equation}
&
\begin{equation}\begin{gathered}
\small \tilde{r}^{-}_{203} = \frac{⟨25⟩⟨13⟩[13]^2}{⟨14⟩⟨15⟩[15]^2⟨34⟩}
\end{gathered}\hspace{-2mm}\nonumber
\end{equation}
&
\begin{equation}\begin{gathered}
\small \tilde{r}^{-}_{223} = \frac{[13]^2⟨12⟩⟨24⟩⟨13⟩}{⟨14⟩^3[15]^2⟨23⟩}
\end{gathered}\hspace{-2mm}\nonumber
\end{equation}
\\
\end{tabular}

\pagebreak

\hskip-0.4cm
\begin{tabular}{p{4.5cm}@{}p{4.5cm}@{}p{4.5cm}@{}p{4.5cm}}
\begin{equation}\begin{gathered}
\small \tilde{r}^{-}_{241} = \frac{⟨34⟩⟨25⟩^2[34]^2}{⟨13⟩⟨24⟩⟨4|1\tiny{+}3|4]^2}
\end{gathered}\hspace{-2mm}\nonumber
\end{equation}
&
\begin{equation}\begin{gathered}
\small \tilde{r}^{-}_{261} = \frac{⟨14⟩[34]^3[13]^2}{[23][35]^2⟨3|1\tiny{+}2|3]^2}
\end{gathered}\hspace{-2mm}\nonumber
\end{equation}
&
\begin{equation}\begin{gathered}
\small \tilde{r}^{-}_{281} = \frac{⟨45⟩⟨35⟩⟨12⟩[14]}{[12]⟨13⟩^2⟨24⟩⟨34⟩}
\end{gathered}\hspace{-2mm}\nonumber
\end{equation}
&
\begin{equation}\begin{gathered}
\small \tilde{r}^{-}_{301} = \frac{⟨25⟩^3⟨34⟩[13]}{⟨13⟩[15]⟨24⟩^2⟨35⟩⟨45⟩}
\end{gathered}\hspace{-2mm}\nonumber
\end{equation}
\\[-9.5mm]
\begin{equation}\begin{gathered}
\small \tilde{r}^{-}_{242} = \frac{[34]^2⟨15⟩[13]^2}{[23][35]⟨3|1\tiny{+}2|3]^2}
\end{gathered}\hspace{-2mm}\nonumber
\end{equation}
&
\begin{equation}\begin{gathered}
\small \tilde{r}^{-}_{262} = \frac{⟨24⟩^3⟨13⟩^3⟨25⟩^2}{⟨12⟩^3⟨14⟩⟨23⟩⟨34⟩^4}
\end{gathered}\hspace{-2mm}\nonumber
\end{equation}
&
\begin{equation}\begin{gathered}
\small \tilde{r}^{-}_{282} = \frac{[34][14]⟨15⟩⟨25⟩}{⟨14⟩⟨23⟩⟨24⟩[24]^2}
\end{gathered}\hspace{-2mm}\nonumber
\end{equation}
&
\begin{equation}\begin{gathered}
\small \tilde{r}^{-}_{302} = \frac{⟨45⟩⟨12⟩^2⟨15⟩[13]}{⟨14⟩^3⟨23⟩⟨35⟩[35]}
\end{gathered}\hspace{-2mm}\nonumber
\end{equation}
\\
\end{tabular}

\pagebreak

\hskip-0.7cm
\begin{tabular}{p{4.6cm}@{}p{4.7cm}@{}p{4.7cm}@{}p{4.7cm}}
\begin{equation}\begin{gathered}
\small \tilde{r}^{-}_{321} \kern-1mm = \kern-1mm \frac{⟨14⟩[34]^2⟨25⟩^2}{⟨13⟩⟨15⟩⟨24⟩⟨35⟩[35]^2}
\end{gathered}\hspace{-2mm}\nonumber
\end{equation}
&
\begin{equation}\begin{gathered}
\small \tilde{r}^{-}_{341}\kern-1mm = \kern-1mm\frac{⟨12⟩[12][34]⟨15⟩}{⟨13⟩⟨14⟩[25]⟨1|2\tiny{+}5|1]}
\end{gathered}\hspace{-2mm}\nonumber
\end{equation}
&
\begin{equation}\begin{gathered}
\small \tilde{r}^{-}_{361}\kern-1mm = \kern-1mm\frac{[12]^2[24]⟨24⟩^2⟨23⟩}{[15][25]⟨34⟩^3⟨2|1\tiny{+}5|2]}
\end{gathered}\hspace{-2mm}\nonumber
\end{equation}
&
\begin{equation}\begin{gathered}
\small \tilde{r}^{-}_{381}\kern-1mm = \kern-1mm\frac{⟨25⟩^2⟨14⟩[34][13]}{⟨13⟩⟨15⟩[15]⟨24⟩[35]⟨45⟩}
\end{gathered}\hspace{-2mm}\nonumber
\end{equation}
\\[-9.5mm]
\begin{equation}\begin{gathered}
\small \tilde{r}^{-}_{322}\kern-1mm = \kern-1mm\frac{⟨35⟩^2[35]^2⟨15⟩}{⟨13⟩⟨14⟩⟨23⟩[25]^2⟨45⟩}
\end{gathered}\hspace{-2mm}\nonumber
\end{equation}
&
\begin{equation}\begin{gathered}
\small \tilde{r}^{-}_{342}\kern-1mm = \kern-1mm\frac{[35][14]⟨15⟩⟨25⟩}{⟨14⟩⟨23⟩[25]⟨5|1\tiny{+}4|5]}
\end{gathered}\hspace{-2mm}\nonumber
\end{equation}
&
\begin{equation}\begin{gathered}
\small \tilde{r}^{-}_{362}\kern-1mm = \kern-1mm\frac{[34]^2⟨13⟩⟨15⟩⟨24⟩^2}{⟨14⟩^3⟨23⟩[45]⟨4|1\tiny{+}2|4]}
\end{gathered}\hspace{-2mm}\nonumber
\end{equation}
&
\begin{equation}\begin{gathered}
\small \tilde{r}^{-}_{382}\kern-1mm = \kern-1mm\frac{⟨24⟩⟨12⟩[34]^2⟨13⟩[13]}{⟨14⟩^3[14]⟨23⟩[35][45]}
\end{gathered}\hspace{-2mm}\nonumber
\end{equation}
\\
\end{tabular}

\pagebreak

\vspace{-2mm}
\hskip-0.75cm
\begin{tabular}{p{4.5cm}@{}p{4.6cm}@{}p{4.7cm}@{}p{4.8cm}}
\begin{equation}\begin{gathered}
\small \tilde{r}^{-}_{398} = \frac{2⟨35⟩^2[13]^3}{⟨12⟩[12]^3⟨24⟩⟨34⟩}\\
\small+\frac{3⟨25⟩⟨35⟩[13]^2}{⟨12⟩[12]^2⟨24⟩⟨34⟩}\phantom{+}
\end{gathered}\hspace{-2mm}\nonumber
\end{equation}
&
\begin{equation}\begin{gathered}
\small \tilde{r}^{-}_{408} = \frac{⟨15⟩[15]⟨45⟩[14]}{[25]⟨34⟩^2⟨35⟩[35]}\\
\small+\frac{-[14]⟨25⟩[24][23]}{[25]^2⟨35⟩⟨2|1\tiny{+}4|2]}\phantom{+}
\end{gathered}\hspace{-2mm}\nonumber
\end{equation}
&
\begin{equation}\begin{gathered}
\small \tilde{r}^{-}_{418} = \frac{⟨45⟩[35]⟨35⟩^2[15]⟨25⟩}{⟨34⟩^3⟨5|1\tiny{+}2|5]^2}\\
\small+\frac{-3⟨45⟩⟨24⟩⟨35⟩^2⟨25⟩[25]}{⟨15⟩⟨34⟩^4⟨5|1\tiny{+}2|5]}\phantom{+}
\end{gathered}\hspace{-2mm}\nonumber
\end{equation}
&
\begin{equation}\begin{gathered}
\small \tilde{r}^{-}_{427} = \frac{-[45]⟨12⟩⟨35⟩⟨15⟩⟨45⟩}{⟨13⟩^2⟨14⟩⟨23⟩[23]⟨34⟩}\\
\small+\frac{[45]⟨12⟩⟨35⟩^2⟨45⟩[24]}{[12]⟨13⟩^3⟨23⟩[23]⟨34⟩}\phantom{+}
\end{gathered}\hspace{-2mm}\nonumber
\end{equation}
\\[-8.5mm]
\begin{equation}\begin{gathered}
\small \tilde{r}^{-}_{399} = \frac{3[15]^2⟨35⟩^2⟨45⟩^2}{[12]⟨34⟩^4⟨5|1\tiny{+}2|5]}\\
\small+\frac{[15][13]⟨35⟩^2}{[12]⟨34⟩^3[45]}\phantom{+}
\end{gathered}\hspace{-2mm}\nonumber
\end{equation}
&
\begin{equation}\begin{gathered}
\small \tilde{r}^{-}_{409} = \frac{[24]^2⟨24⟩^2⟨25⟩^2}{⟨12⟩⟨34⟩^2⟨2|1\tiny{+}5|2]^2}\\
\small+\frac{2[24]⟨24⟩^2⟨35⟩⟨25⟩}{⟨12⟩⟨34⟩^3⟨2|1\tiny{+}5|2]}\phantom{+}
\end{gathered}\hspace{-2mm}\nonumber
\end{equation}
&
\begin{equation}\begin{gathered}
\small \tilde{r}^{-}_{419} = \frac{2⟨35⟩^2[24][23]^2⟨23⟩^2}{⟨12⟩[12]^3⟨13⟩^3⟨34⟩}\\
\small+\frac{-⟨35⟩[24]⟨25⟩⟨23⟩[23]}{⟨12⟩[12]^2⟨13⟩^2⟨34⟩}\phantom{+}
\end{gathered}\hspace{-2mm}\nonumber
\end{equation}
&
\begin{equation}\begin{gathered}
\small \tilde{r}^{-}_{428} = \frac{[24]⟨24⟩[12]⟨25⟩^2}{⟨23⟩[25]⟨34⟩⟨45⟩⟨5|1\tiny{+}3|5]}\\
\small+\frac{-⟨25⟩[13]^2⟨12⟩⟨35⟩}{⟨23⟩⟨34⟩[35]⟨45⟩⟨5|1\tiny{+}3|5]}\phantom{+}
\end{gathered}\hspace{-2mm}\nonumber
\end{equation}
\\
\end{tabular}

\pagebreak

\vspace{-2mm}
\hskip+0cm
\begin{tabular}{p{5.2cm}@{}p{5.8cm}@{}p{5.8cm}@{}}
\begin{equation}\begin{gathered}
\small \tilde{r}^{-}_{435} = \frac{2⟨12⟩[45]⟨15⟩^2}{⟨13⟩^3⟨14⟩[35]}+\\
\small\frac{-⟨12⟩[45]^2⟨24⟩⟨15⟩^2}{⟨13⟩^3⟨14⟩⟨23⟩[35]^2}+\\
\small\frac{2[34][15]⟨15⟩⟨12⟩^2}{⟨13⟩^2⟨14⟩⟨23⟩[35]^2}\phantom{+}
\end{gathered}\hspace{-2mm}\nonumber
\end{equation}
&
\begin{equation}\begin{gathered}
\small \tilde{r}^{-}_{441} = \frac{3⟨13⟩[24]⟨25⟩⟨24⟩^2⟨23⟩}{⟨12⟩^2⟨34⟩^4[45]}+\\
\small\frac{[14]⟨13⟩^2[34]⟨24⟩^2⟨23⟩}{⟨12⟩^2⟨34⟩^4[45]^2}+\\
\small\frac{2⟨13⟩⟨25⟩[34]⟨24⟩⟨23⟩}{⟨12⟩^2⟨34⟩^3[45]}\phantom{+}
\end{gathered}\hspace{-2mm}\nonumber
\end{equation}
&
\begin{equation}\begin{gathered}
\small \tilde{r}^{-}_{447} = \frac{2[14]^2⟨25⟩[13]⟨15⟩}{[15]⟨35⟩⟨1|2\tiny{+}4|1]^2}+\\
\small\frac{2⟨12⟩[13]^2[34][14]}{[15][35]⟨1|2\tiny{+}4|1]^2}+\\
\small\frac{[14][13][34]^2}{[15][24][35]⟨1|2\tiny{+}4|1]}+\\
\small\frac{-[14]^2[34]⟨15⟩}{⟨13⟩[15][24]⟨1|2\tiny{+}4|1]}+\\
\small\frac{[14]⟨25⟩^2[13]}{[15]⟨24⟩⟨35⟩⟨1|2\tiny{+}4|1]}\phantom{+}
\end{gathered}\hspace{-2mm}\nonumber
\end{equation}
\\
\end{tabular}

\twocolumngrid
\bibliography{main_quarks}

\end{document}